\newcommand{\PRE}[1]{{#1}} 
\documentclass[11pt,tightenlines,showpacs,preprintnumbers,%
amsmath,amssymb,preprint,superscriptaddress,nofootinbib]{revtex4}

\usepackage{enumerate}
\usepackage{color}
\usepackage[pdftex]{graphicx}
\usepackage[colorlinks=true,citecolor=blue,urlcolor=blue]{hyperref}
\graphicspath{ {../Figures/} }

\newcommand{\mplanck}{M_{\text{Pl}}}

\newcommand{\mev}{\text{MeV}}
\newcommand{\gev}{\text{GeV}}
\newcommand{\tev}{\text{TeV}}

\newcommand{\cm}{\text{cm}}

\newcommand{\km}{\text{km}}
\newcommand{\g}{\text{g}}
\newcommand{\s}{\text{s}}

\renewcommand{\eqref}[1]{Eq.~(\ref{#1})}

\newcommand{\secref}[1]{Sec.~\ref{sec:#1}}
\newcommand{\secsref}[2]{Secs.~\ref{sec:#1} and \ref{sec:#2}}

\newcommand{\figref}[1]{Fig.~\ref{fig:#1}}
\newcommand{\figsref}[2]{Figs.~\ref{fig:#1} and \ref{fig:#2}}
\newcommand{\Figref}[1]{Figure~\ref{fig:#1}}

\newcommand{\order}[1]{\mathcal{O}(#1)}
\newcommand{\units}[1]{\mathrm{#1}}
\def\Neff{N_\textrm{eff}}
\newcommand{\gb}{\text{gb}}
\newcommand{\gbino}{\text{gbino}}

\begin{document}

\preprint{CALT-68-2860}
\preprint{UCI-TR-2013-18}

\title{\PRE{\vspace*{1.5in}}
Self-interacting dark matter from a non-Abelian hidden sector
\PRE{\vspace*{.5in}}}

\author{Kimberly K.~Boddy}
\affiliation{California Institute of Technology, Pasadena, California
  91125, USA}

\author{Jonathan L.~Feng}
\author{Manoj Kaplinghat}
\author{Tim M.~P.~Tait\PRE{\vspace*{.1in}}}
\affiliation{Department of Physics and Astronomy, University of California, Irvine, California 92697, USA
\PRE{\vspace*{.5in}}}

\begin{abstract}

\PRE{\vspace*{.2in}}
There is strong evidence in favor of the idea that dark matter is
self interacting, with the cross section-to-mass ratio $\sigma / m \sim
1~\cm^2/\g \sim 1~\text{barn}/\gev$.  We show that viable models of
dark matter with this large cross section are straightforwardly
realized with non-Abelian hidden sectors.  In the simplest of such
models, the hidden sector is a pure gauge theory, and the dark matter
is composed of hidden glueballs with a mass around 100 MeV.  Alternatively, the
hidden sector may be a supersymmetric pure gauge theory with a $\sim
10~\tev$ gluino thermal relic.  In this case, the dark matter is
largely composed of glueballinos that strongly self interact through
the exchange of light glueballs.  We present a unified framework that
realizes both of these possibilities in anomaly-mediated supersymmetry
breaking, where, depending on a few model parameters, the dark matter
may be composed of hidden glueballinos, hidden glueballs, or a mixture of the
two.  These models provide simple examples of multicomponent dark
matter, have interesting implications for particle physics and
cosmology, and include cases where a subdominant component of dark
matter may be extremely strongly self interacting, with interesting
astrophysical consequences.

\end{abstract}


\maketitle

\section{Introduction}

The standard model of cosmology describes a universe that is dominated
by the vacuum energy $\Lambda$ and collisionless cold dark matter (CDM).
The success of the $\Lambda$CDM model is based on its well-established
record of describing the features of the large-scale structure
observed in the Universe.

On smaller scales, however, the picture is much less clear.  $N$-body
simulations of collisionless CDM appear to disagree with observations
on small scales, motivating dark matter properties that differ
significantly from the standard paradigm.  In particular, if dark
matter is self interacting (able to scatter elastically with itself),
simulations show that the core sizes and central densities of dwarf
spheroidal galaxies, low-surface-brightness spirals, and galaxy
clusters can all be brought in line with
observations~\cite{Rocha:2012jg,Peter:2012jh,Vogelsberger:2012ku,Zavala:2012us}.
Modifying $\Lambda$CDM to incorporate self-interacting dark matter
(SIDM), sometimes called the $\Lambda$SIDM model, is
consistent with constraints from the Bullet Cluster, measurements of
dark matter halo shapes, and subhalo survival requirements.  To make
the simulations and observations consistent, the ratio of the dark
matter's self-interaction cross section to its mass should be in the
range $\sigma/m \sim 0.1 - 10~\units{cm^2/g}$. The requirement of such
strong self interactions eliminates from consideration all of the most
studied dark matter candidates, including weakly interacting massive
particles (WIMPs), axions, and sterile neutrinos.

At the same time, such large cross sections are on par with
nuclear-scale cross sections ($1~\units{cm^2/g} \simeq
1.78~\units{barn/GeV}$).  The possibility that dark matter has color
and interacts through the strong interactions of the standard model
(SM) is highly constrained, for example, by searches for anomalous
isotopes in sea
water~\cite{Hemmick:1989ns,Verkerk:1991jf,Yamagata:1993jq}.  However,
dark matter may self interact through non-Abelian forces (such as a
dark analogue of QCD) in a hidden sector.  As we will show below, this
is straightforwardly realized in even the simplest such hidden
sectors, with SU($N$) gauge symmetry and no additional matter content.
For the confinement scales $\Lambda \sim 100~\mev$, the hidden gluons
confine to form glueballs, and the resulting glueball dark matter has
the required self interactions.  For hidden sectors that are roughly
the same temperature as the visible sector, the glueball relic density
is generically too large, but the desired relic density may be
obtained by adjusting the relative temperatures of the hidden sector
and visible sector, as we discuss below.

This hidden glueball scenario for self-interacting dark matter is
remarkably simple, but it is decoupled from the visible sector, both
in the technical sense and in the sense that it is not motivated by
any of the well-known problems of the SM.  In addition, the correct
relic density is arranged by tuning a free parameter, the ratio of
hidden to visible sector temperatures, and so the scenario cannot be
claimed to naturally produce the right thermal relic density, as in
the case of WIMPs.  At first sight, it might appear to be difficult to
enhance the model to accommodate all of these desirable features,
especially since the WIMP miracle requires weak-scale annihilation
cross sections, whereas the required self interactions naturally
suggest strong interactions.

In fact, however, we will show that all of these features are present
in a supersymmetric version of the hidden glueball scenario, in which
the hidden sector is a supersymmetric pure gauge theory.  In this
model, the dark matter is a $\sim 10~\tev$ hidden gluino, which
freezes out in the early Universe when the temperature is high.  At
freeze-out, the theory is weakly coupled, but, as the Universe cools and
expands, the theory confines, forming hidden glueballinos and
glueballs.  The glueballinos strongly interact via exchange of the
hidden glueballs with the required self-interaction cross section.
This scenario is straightforwardly accommodated in anomaly-mediated
supersymmetry breaking (AMSB)
scenarios~\cite{Randall:1998uk,Giudice:1998xp}, which provide a
connection to the problems of the SM and also allow the glueballinos
to naturally inherit the correct relic density through the WIMPless
miracle~\cite{Feng:2008ya,Feng:2008mu}, a possibility discussed
previously in Refs.~\cite{Feng:2011ik,Feng:2011uf,Feng:2011in}.  For
related work on strongly interacting hidden sectors and dark matter,
see Refs.~\cite{Kribs:2009fy,Alves:2009nf,Falkowski:2009yz,%
  Lisanti:2009am,Alves:2010dd,Kumar:2011iy,Cline:2013zca,Kang:2006yd,%
  Higaki:2013vuv,Heikinheimo:2013fta,Bai:2013xga}.

\begin{figure}[t]
  \centering
  \includegraphics[scale=0.7]{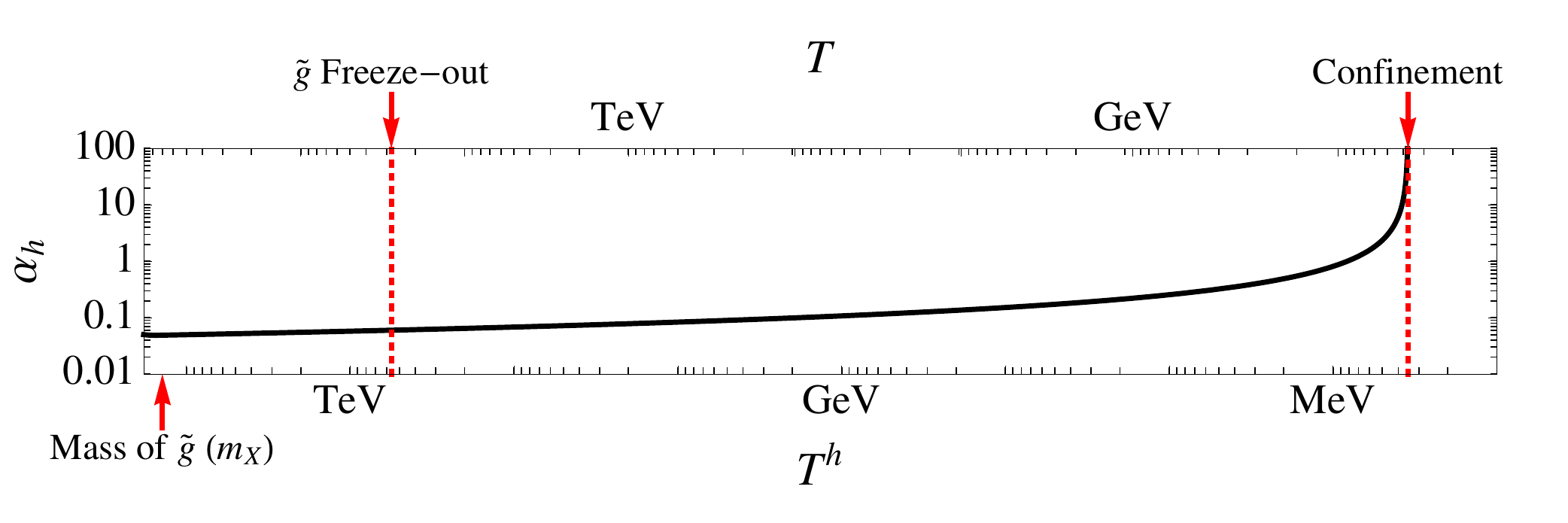}
  \caption{Example timeline of events in the supersymmetric pure
    SU($N$) theory without connectors, in terms of the hidden- and
    visible-sector temperatures $T^h$ and $T$.  The hidden-sector
    coupling $\alpha_h$ is shown as a function of these temperatures.
    It is weak at gluino freeze-out but grows as the temperature drops,
    leading to confinement and the formation of glueballino and
    glueball dark matter at a temperature $\sim \Lambda$. The scenario
    is described in detail in \secref{noconnectors}, and the chosen
    parameters are represented by the yellow dot in
    \figref{AMSB-pure-relic}.  }
  \label{fig:timeline1}
\end{figure}

Of course, the supersymmetric models also contain glueballs, which, as
in the nonsupersymmetric case, may be dark matter.  As we will see,
in different regions of the AMSB parameter space, the dark matter may
be composed of mostly glueballinos, mostly glueballs, or a mixture of the
two.  For the case where the dark matter is composed of mostly glueballinos,
we detail two possibilities. In the first case, the hidden sector is
coupled to the visible sector only indirectly through the
supersymmetry breaking mechanism. Since this coupling is extremely
weak, the sectors can have different temperatures, and the glueball
relic density may be very small for cold hidden sectors. An example
cosmological timeline of events in this case is given in
\figref{timeline1}.

\begin{figure}[t]
  \centering
  \includegraphics[scale=0.7]{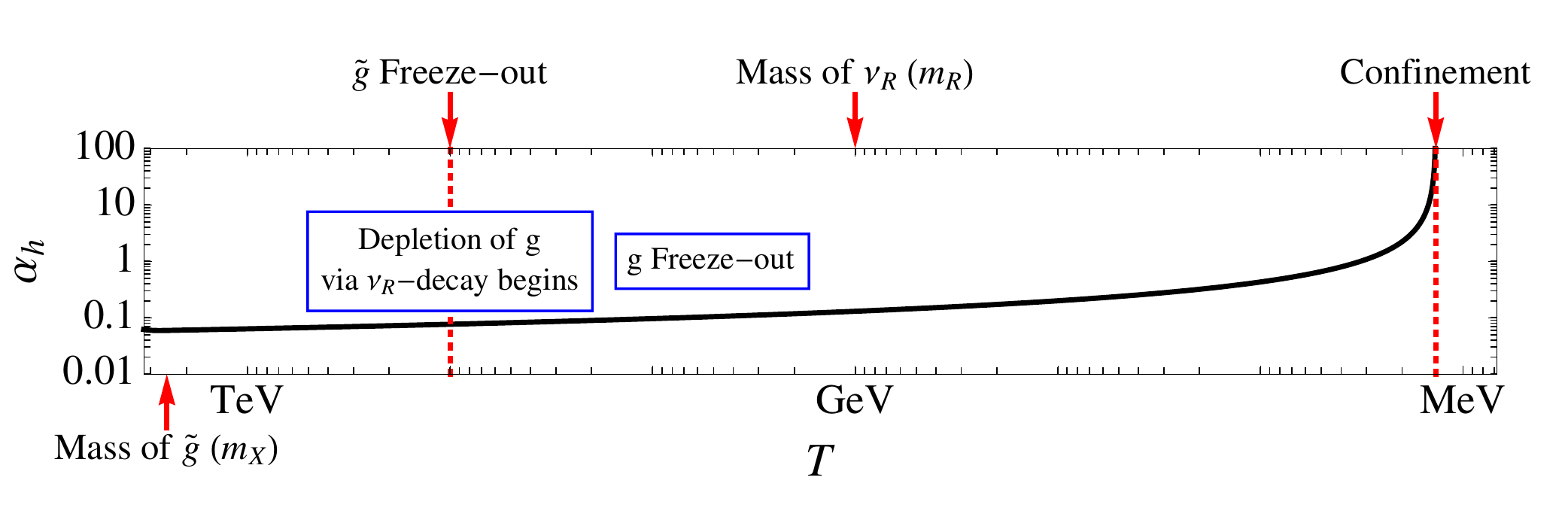}
  \caption{As in \figref{timeline1}, but for supersymmetric pure
      SU($N$) theory {\em with} connectors.  Since the hidden and
      visible sectors communicate efficiently in the early Universe,
      they share a temperature, $T$.  The gluon population is
      depleted through annihilations to and the subsequent decays of
      the $\nu_R$ in the visible sector, and the resulting scenario
      has pure glueballino dark matter.  The scenario is described in
      detail in \secref{connectors}, and the chosen parameters are
      represented by the yellow dot in \figref{AMSB-pureConn-relic}. }
  \label{fig:timeline2}
\end{figure}

In the second case, the hidden sector is coupled to the visible sector
through connector fields.  The visible and hidden sectors, therefore,
have the same temperature at early times, leading, {\em a priori}, to
too-large glueball relic densities.  Decays of glueballs are in
conflict with big bang nucleosynthesis (BBN) and other astrophysical
and particle constraints.  Instead, we rely on a novel nonthermal
process in the early Universe to deplete the gluon density, thereby
suppressing the glueball density after confinement.  In this case, the
gluons annihilate into singlet right-handed neutrinos with $\sim
1~\units{GeV}$ mass, and we reduce the hidden gluon density by forcing
the right-handed neutrinos to decay into SM particles more quickly
than they can annihilate back into hidden gluons.  A representative
timeline for this case is shown in \figref{timeline2}.

This work is organized as follows. In \secref{evidence}, we review the
astrophysical evidence for self-interacting dark matter.  In
\secref{glueball}, we begin with the simplest possible case:
nonsupersymmetric pure gauge hidden sectors and glueball dark matter.
We discuss glueball self interactions and relic densities and
determine the preferred parameters for this simple model.  We then
move to supersymmetric models with pure gauge hidden sectors and
glueballino dark matter.  In \secref{glueballinoscattering}, we review
the calculation of the glueballino self-interaction cross section, and
in \secref{glueballinorelic}, we discuss the glueballino relic density
and the realization of the WIMPless miracle in the AMSB framework.
Finally, with this groundwork, we present full AMSB models of
glueballino/glueball dark matter without and with connectors in
\secsref{noconnectors}{connectors}, respectively. We conclude in
\secref{conclusions}.

Last, we make a note on naming conventions. In the rest of this work, we
follow the literature: glueballinos denote gluino-gluon bound states,
while gluinoballs denote gluino-gluino bound states.  In addition,
unless otherwise stated, gluon, gluino, glueball, and
glueballino refer to hidden sector particles and are denoted by
$g$, $\tilde{g}$, $\gb$, and $\gbino$, respectively.

\section{Astrophysical evidence for self-interacting dark matter}
\label{sec:evidence}

The $\Lambda$CDM model is quite successful in describing the large-scale
structure.  The predictions of the standard six-parameter $\Lambda$CDM
cosmology match remarkably well to the latest measurements of the
cosmic microwave background (CMB) by the WMAP~\cite{Hinshaw:2012aka} and
Planck~\cite{Ade:2013zuv} at large multipoles of the power spectrum.
Additionally, CDM fits the dark matter power spectrum very
well~\cite{Reid:2009xm}, using observations of luminous red galaxy
clustering in the Sloan Digital Sky Survey~\cite{Abazajian:2008wr}.

Despite these agreements on large scales, observations of small-scale
structures indicate that CDM is insufficient.  Challenges to the
$\Lambda$CDM paradigm arise largely from tensions between observation
and cosmological simulations.  Simulations of CDM create dark matter
halos with density profiles that have steep, inverse-power-law
behaviors (cusps) towards the center of the
halo~\cite{Navarro:1994hg,Navarro:1995iw,Navarro:1996gj,%
  Bullock:1999he,Wechsler:2001cs}.  Conversely, observations show that
low-surface-brightness spiral galaxies (LSBs)~\cite{Flores:1994gz,%
  Simon:2004sr,KuziodeNaray:2007qi,Dutton:2010dw,deNaray:2011hy,%
  Oh:2010ea,Salucci:2011ee,Castignani:2012sr}, satellite dwarf
galaxies~\cite{Walker:2011zu,Salucci:2011ee}, and galaxy
clusters~\cite{Sand:2003ng,Sand:2007hm,Newman:2009qm,Newman:2011ip,%
  Coe:2012kj,Umetsu:2012nj} exhibit constant-density cores.  In
addition to the core-cusp discrepancy~\cite{deBlok:2009sp}, the
simulated central densities of halos are too high.  By matching the
luminosity function of the Milky Way to the Aquarius
simulations~\cite{Springel:2008cc}, the brightest subhalos in the
Milky Way are a factor of 5 less massive than
predicted~\cite{BoylanKolchin:2011de,BoylanKolchin:2011dk}.  If
$\Lambda$CDM is correct, we are left to explain this
``too-big-to-fail" problem in which the largest subhalos of the Milky
Way do not luminesce; otherwise, some additional physics is needed in
simulations to decrease the central densities of these overly massive
halos.

To address these concerns with $\Lambda$CDM, there are a few generic
possibilities to consider~\cite{Weinberg:2013aya}: adding feedback
from baryons in
simulations~\cite{Oh:2010mc,Governato:2012fa,Newman:2012nw}, warm dark
matter (WDM)~\cite{Tremaine:1979we,Bond:1980ha,Olive:1981ak}, and
self-interacting dark
matter~\cite{Spergel:1999mh,Firmani:2000ce,Firmani:2000qe}.  Feedback
exists and should be included in simulations, but there may not be
enough energy to eject a sufficient amount of mass from the halo
center to solve the too-big-to-fail
problem~\cite{BoylanKolchin:2011dk}.  WDM tends to be too efficient in
wiping out structure, leaving too few subhalos in the Milky
Way~\cite{Polisensky:2010rw}.  Additionally, lower bounds on WDM
masses from Lyman-$\alpha$ forest measurements constrain the ability
of WDM to solve the core-cusp problem over the full range of
astrophysical scales needed~\cite{Seljak:2006qw,Viel:2013fqw}.  Even
with its mass unconstrained, WDM still leaves dwarf halos cuspy,
though it does lower the central densities~\cite{deNaray:2009xj}.

On the other hand, self-interacting dark matter can soften halo cores
and lower central densities, while preserving large-scale
structure~\cite{Spergel:1999mh} and satisfying bounds of $\sigma/m
\lesssim 1~\units{cm^2/g}$ from the Bullet
Cluster~\cite{Randall:2007ph}.  Indeed, simulations with constant dark
matter cross section-to-mass ratios in the range $0.1 -
1~\units{cm^2/g}$ show that self interactions can bring theory in line
with observations of both halo profiles and
shapes~\cite{Rocha:2012jg,Peter:2012jh}.  Velocity-dependent
self interactions widen this range to $0.1 - 10~\units{cm^2/g}$ and
can also soften cores and reduce the density of the brightest
satellites to solve the too-big-to-fail
problem~\cite{Vogelsberger:2012ku,Zavala:2012us}.

With these results from simulation, dark matter with self interactions
is a strong contender within particle physics to be a solution to the
small-scale formation woes in astrophysics.  From a particle physics
perspective, we will see that self scattering is a quite reasonable
and perhaps even generic property for dark matter to possess.

\section{Glueball dark matter}
\label{sec:glueball}

The simplest construction resulting in dark matter that is a composite
of a strongly interacting hidden sector is a pure Yang-Mills gauge
theory.  At large-energy scales, the theory consists of a weakly
coupled set of massless gluons whose couplings are described by
the gauge coupling.  The theory is expected to confine at low energies
at a scale $\Lambda$, where the gauge coupling becomes strong enough
that perturbation theory breaks down~\cite{Dietrich:2006cm,%
  Appelquist:1998rb,Miransky:1996pd,Cohen:1988sq,Appelquist:1988yc,%
  Poppitz:2009tw,Poppitz:2009uq}.  At this point it develops a mass
gap, and the low energy physics is described by a set of glueball
states whose masses are characterized by $\Lambda$ through
dimensional transmutation.

At very low energies $\ll \Lambda$, the physics is described by an
effective field theory composed of the low-lying glueball states.
In the absence of any coupling to the SM, the lightest of these states
will be effectively stable.\footnote{Note that gravitational
  interactions will mediate very suppressed decays to light SM
  particles, but these are irrelevant for $\Lambda \ll \mplanck$.}
The detailed mass spectrum (and spins) of these states depends on the
underlying choice of theory and is further clouded by strong coupling,
which leaves results based on perturbation theory suspect.
Generically, one expects the glueball spectrum to have a lowest-lying
element whose mass is ${\cal O}(\Lambda)$, which, following the
guidance of QCD, we take to be a $J^{CP}=0^{++}$
state~\cite{Narison:1997nw,Morningstar:1999rf}.  There will also be a
collection of excited states with various spins and whose mass
splittings are roughly multiples of $\Lambda$.

\subsection{Glueball self interactions}
\label{sec:glueballscattering}

The various glueball states will interact with one another as a
residual of the strong dynamics that bind them.  Dimensional analysis
dictates that the interactions among them will be proportional to
$\Lambda$ to an appropriate power, with dimensionless coefficients
characterized by na\"{i}ve dimensional analysis (NDA)
\cite{Manohar:1983md,Cohen:1997rt}.  For example, a description of a
scalar glueball state, $\phi_0$, would look like
\begin{eqnarray}
{\cal L}_{\gb} & = &
\frac{1}{2} \partial_\mu \phi_0 \partial^\mu \phi_0 - \frac{1}{2} m^2 \phi_0^2
+ \frac{A}{3!} \phi_0^3 + \frac{\lambda}{4!} \phi_0^4 + \ldots \ ,
\end{eqnarray}
where NDA would suggest that for the lowest-lying state $m \simeq
\Lambda$, and $A \simeq (4 \pi) \Lambda$, $\lambda \simeq (4 \pi)^2$,
and the $+...$ indicates interaction terms in the form of higher
dimensional operators that are suppressed by powers of $\Lambda$.
Interactions involving the various glueball excited states can be
formulated in a similar way.

For energies $\ll \Lambda$, the physics should be well described by an
effective field theory composed of the lightest glueball.  At kinetic
energies of the order $\Lambda$, more of the lowest lying states become
accessible and need to be included in the effective theory.  At
energies $\gg \Lambda$, the physics is described by the interactions
of the gluons together with the structure functions that describe
their distribution inside of the glueballs.

Although it is clear that glueballs are strongly self interacting, it
is very difficult to make precise predictions for the scattering rate,
given our general ignorance concerning strongly coupled theories.  The
expected cross section will be characterized by the confinement scale
and strong coupling,
\begin{eqnarray}
\sigma \left( \gb~\gb \rightarrow \gb~\gb \right) \simeq \frac{4 \pi}{\Lambda^2}~,
\end{eqnarray}
which can also be understood from the geometric size of the glueballs,
whose radius is $\sim 1 / \Lambda$.

\subsection{Glueball relic density}
\label{sec:glueballrelic}

If the glueballs are stable on the scale of the age of the Universe,
they will contribute to the total observed dark matter relic density.
At early times, when their temperature is $T_h \gg \Lambda$, the
hidden sector is represented by a plasma of gluons whose comoving
relativistic number density is given by
\begin{equation}
  Y_\infty = \frac{n_{g}}{s}
  = \left. \frac{[\zeta(3)/\pi^2]g_\textrm{eff} {T^h}^3}{(2\pi^2/45)g_{*S}T^3}
  \right|_{t_f}
  = \frac{45\zeta(3)}{2\pi^4} \xi_f^3 \frac{g_\textrm{eff}}{g_{*S}(t_f)} \ ,
  \label{eq:Yinf}
\end{equation}
where $s$ is the entropy in the visible sector, $\xi_f \equiv T^h / T$
is the ratio of temperatures in the hidden and visible sectors,
$\zeta(3)\approx 1.202$ is the zeta function, and
$g_\textrm{eff}=2(N^2-1)$ for an SU($N$) gauge theory.  We use an
early time, $t_f$ (which we will identify with the time of gluino
freeze-out in the supersymmetric models discussed below), as a reference
point.  The quantity $Y_\infty$ remains constant as the Universe
expands.

As the hidden sector temperature $T^h$ cools below the critical
temperature $T^c \sim \Lambda$ \cite{Lucini:2005vg}, there is a
transition to the confined phase, and the energy density of the gluon
plasma is converted into glueballs.  The result is that after
confinement the Universe is filled with nonrelativistic glueballs
whose comoving number density is the same as that of the gluons up to
factors of $\order{1}$.  Consequently, today the glueballs are
nonrelativistic with a relic density:
\begin{equation}
  \Omega_{\gb} \sim \frac{Y_\infty s_0 \Lambda}{\rho_{c0}} \ .
  \label{eq:gbRelic}
\end{equation}
This expression assumes that there are no number-changing processes,
but the glueballs may interact through a dimension-5 operator to give
$3 \to 2$ scatterings~\cite{Carlson:1992fn}. We ignore this
possibility here and leave it for future work.

\subsection{Viable glueball parameters}
\label{sec:glueballmodels}

\begin{figure}[t]
  \centering
  \includegraphics[scale=0.7]{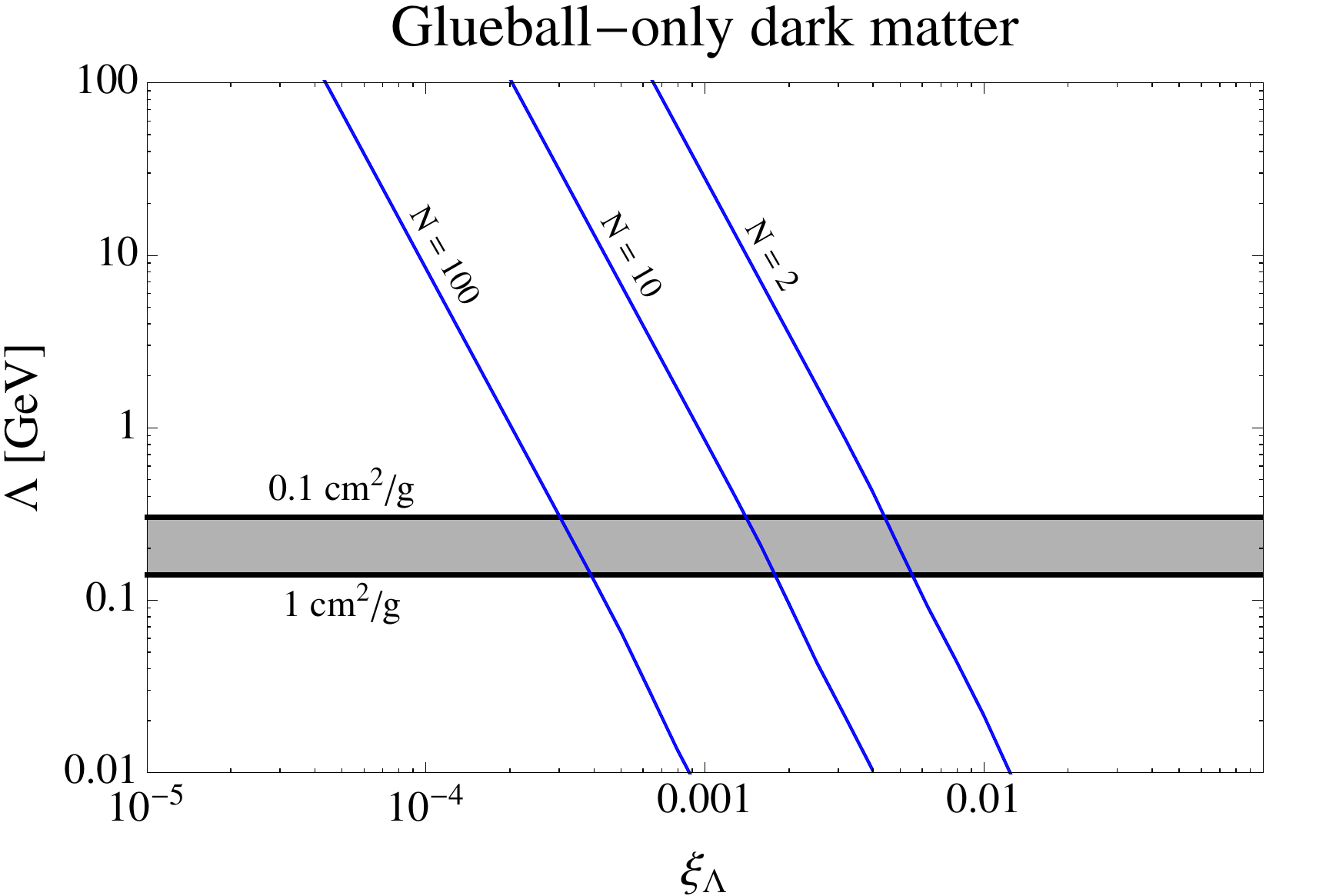}
  \caption{Glueball dark matter in the case of a
    nonsupersymmetric pure gauge SU($N$) hidden sector. The
    self-interaction cross section and relic density are given in the
    $(\xi_{\Lambda}, \Lambda)$ plane, where $\Lambda$ is the
    confinement scale in the hidden sector, and $\xi_{\Lambda} \equiv
    T^h/T$ is the ratio of hidden to visible sector temperatures at
    the time that $T^h = \Lambda$. The self-interaction cross section
    is in the range $\langle \sigma_T \rangle / m_X=0.1 -
    1~\units{cm^2/g}$ in the shaded region.  The glueball relic
    density is $\Omega_{\gb} = \Omega_{\text{DM}} \simeq 0.23$ on the
    diagonal contours for the number of colors $N$ indicated.
  \label{fig:glueball}}
\end{figure}

Glueball dark matter is thus primarily characterized by two
quantities: the confinement scale $\Lambda$, which simultaneously
controls the dark matter mass and its self-interaction cross section,
and $\xi_\Lambda$, the ratio of temperatures of the hidden and visible
sectors at the time of confinement.  Also relevant is the number of
gluon degrees of freedom; for an SU($N$) gauge theory this is
specified by $N$ through $g_\textrm{eff}=2(N^2-1)$.  In
\figref{glueball}, we show the parameter space in the $(\xi_{\Lambda},
\Lambda)$ plane.  The scattering cross section is independent of
$\xi_\Lambda$, which together with the choice of $N$ controls the
relic density of glueballs.  The scattering cross sections of interest
suggest $\Lambda \sim 100$~MeV, amusingly close to $\Lambda_{\rm QCD}
\approx 300$~MeV.  Note that since the cross section is constant, the
acceptable upper limit from simulations is $1~\units{cm^2/g}$, in
particular, to stay within cluster constraints. This limit will
increase to $10~\units{cm^2/g}$ for velocity-dependent cross sections,
which we begin discussing in \secref{glueballinoscattering}.  The
relic density requires the hidden-sector temperature to be a few
orders of magnitude below the visible temperature at the time of
confinement.

\section{Glueballino self interactions}
\label{sec:glueballinoscattering}

The simplest extension to the pure gauge hidden sector discussed in
\secref{glueball} is to add a massive (mass $m_X \gg \Lambda$) gauge
adjoint Majorana fermion to the theory, resulting in a spectrum with two
types of composites: the bosonic glueballs with a mass $\sim \Lambda$ and
the fermionic states with masses $\sim
m_X$~\cite{Farrar:1995ew,Raby:1997pb,Raby:1998xr,Kauth:2009ud}.  Each
sector contains excited states whose mass splittings are again
characterized by $\Lambda$.  In the absence of further ingredients,
the massive fermionic states are stable because of Lorentz invariance,
and this construction allows one to realize a situation where the dark
matter is (mostly) composed of the heavy composite fermions that
self interact via exchange of the much lighter glueballs, naturally
realizing two widely separated energy scales.  This dark sector is
identical to a softly broken $N=1$ supersymmetric gauge theory and can
be considered the supersymmetric version of the model of
\secref{glueball}.  In that language, the composite fermions are
glueballino states.

The self interactions of glueballinos are dominated by the exchange of
glueballs.  At low energies, when the kinetic energy available is
$\lesssim \Lambda$, the scattering will be elastic.  If there is
sufficient kinetic energy,
\begin{eqnarray}\label{eq:inelastic}
\frac{1}{2} m_X v^2 \geq \Lambda \ ,
\end{eqnarray}
inelastic scattering into excited states and glueball emission becomes
possible, leading to novel effects, such as additional rapid halo
cooling.  The inelastic effects are not modeled in the $\Lambda$SIDM
simulations and so are not well understood.  For the remainder of this
work, we focus on the elastic scattering regime and comment later in
this section on systems where this approximation breaks down.

NDA suggests that the coupling between glueballs and glueballinos is
$\alpha \sim 1$.  Even for elastic scattering, there will be a large
number of distinct glueball states, which are capable of mediating
self interactions of the glueballinos, but the dominant contribution
arises from the lightest glueball states, which mediate the longest
range interactions. Thus, we model the induced potential between two
glueballinos as an attractive Yukawa interaction with a range $\Lambda$
and strength $\alpha \sim 1$:
\begin{equation}
  V(r)=-\frac{\alpha}{r}\exp(-\Lambda r) \ .
\end{equation}

It is common to use the transfer cross section
\begin{equation}
  \sigma_T = \int d\Omega (1-\cos\theta) \frac{d\sigma}{d\Omega}
\end{equation}
to compare predictions to observations and simulations.  We have
numerically solved the Schr\"{o}dinger equation to calculate
$\sigma_T$, following the methods of Ref.~\cite{Tulin:2013teo}.  For
the astrophysical systems of interest, to achieve the desired cross
sections of $0.1 - 10 ~\units{cm^2/g}$ with $m_X \gtrsim \units{TeV}$,
the parameters must be in the classical scattering regime, $m_X v \gg
\Lambda$.  Scattering from Yukawa potentials has been studied in this
regime in the context of classical, complex
plasmas~\cite{Khrapak2003:yukawa,Khrapak2004:yukawa,Khrapak:2004transfer},
and simple analytic fits to numerical results for the transfer cross
section have been derived.  These plasma physics results may be
applied directly to the present dark matter case~\cite{Feng:2009hw}
(in fact, they describe the dark matter model exactly, whereas the
Yukawa potentials are only an approximation to screened Coulomb
interactions in the plasma physics context), and we have checked that
these agree well with our numerical results.

Within a galactic halo or cluster, the dark matter particles have a
velocity distribution that we take to be Maxwell-Boltzmann, and so
\begin{equation}
  f(v_i) = \left(\pi v_0^2 \right)^{-3/2} e^{-v_i^2/v_0^2} \ ,
\end{equation}
where $v_0$ is the mode and $\langle v^2_i \rangle =(3/2)v_0^2$ is the
square of the three-dimensional velocity dispersion.  This
distribution is expected for cross sections of
$\sigma/m=1.0~\units{cm^2/g}$ and above~\cite{Vogelsberger:2012sa};
for the slightly lower cross sections that are still of interest to
us, the distribution may be modified, but we do not expect this to
impact our results significantly. Simulations~\cite{Rocha:2012jg} show
that $\langle v^2_i \rangle \approx (1.2\, V_\textrm{max})^2$, where
$V_\textrm{max}$ is the peak circular velocity of a given system, and
thus $v_0 \approx 0.98\, V_\textrm{max}$.  The astrophysical systems
of interest have values of $V_\textrm{max}$ in the ranges $20 -
50~\units{km/s}$ for dwarfs, $50-130~\units{km/s}$ for LSBs, and $700-
1000~\units{km/s}$ for clusters.  We make a simplistic estimate for
the dark matter escape velocity, $v_\textrm{esc}^2=2v_0^2$, so that
the largest relative velocity between particles is $2\sqrt{2}v_0$.
For two scattering dark matter particles with the velocities $\vec{v}_1$ and
$\vec{v}_2$, the velocity-averaged transfer cross section is
\begin{eqnarray}
  \langle \sigma_T \rangle
  &=& \int \frac{d^3v_1\ d^3v_2}{(\pi v_0^2)^3}
  e^{-v_1^2/v_0^2} e^{-v_2^2/v_0^2} \sigma_T(|\vec{v}_1-\vec{v}_2|) \nonumber\\
  &=& \int_0^{2\sqrt{2}v_0} \frac{d^3v}{(2\pi v_0^2)^{3/2}}
      e^{-v^2/2v_0^2} \sigma_T(v) \ .
  \label{eq:sigmaT-avg}
\end{eqnarray}
Note that although the escape velocity may be an underestimate here,
increasing it by a factor of 10 changes $\langle \sigma_T \rangle $
only at the 1\% level.

\begin{figure}[t]
\centering
\includegraphics[scale=0.7]{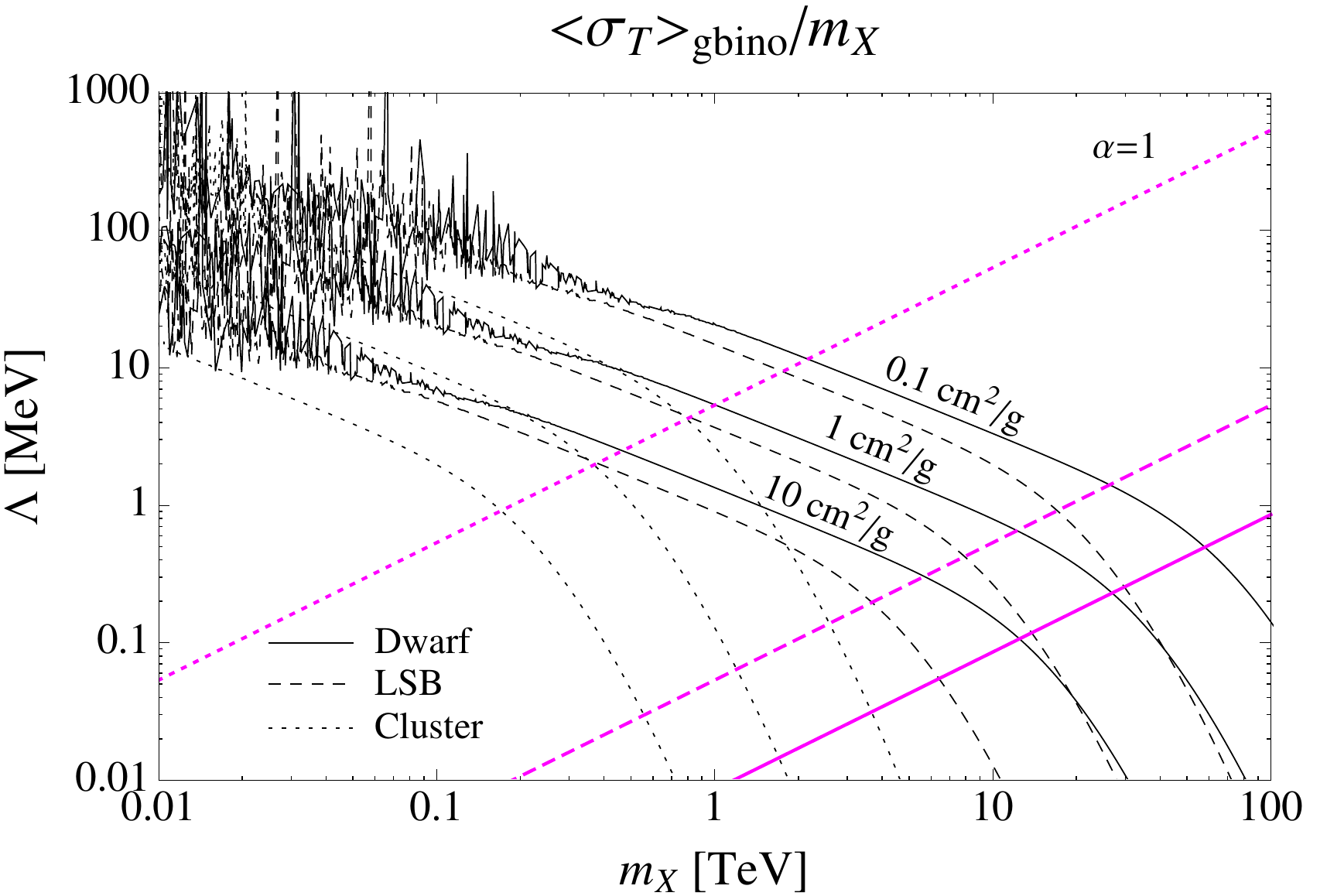}
\caption{The ratio of the thermally averaged transfer cross section to
  the dark matter mass $\langle \sigma_T \rangle /m_X$ in the $(m_X,
  \Lambda)$ plane for $\alpha = 1$ and three different astrophysical
  systems: dwarf galaxies ($V_\textrm{max} = 40~\units{km/s}$, solid),
  LSBs ($V_\textrm{max} = 100~\units{km/s}$, dashed), and clusters
  ($V_\textrm{max} = 1000~\units{km/s}$, dotted).  For each system,
  three values of the cross section are shown: $0.1~\units{cm^2/g}$
  (top), $1~\units{cm^2/g}$ (middle), and $10~\units{cm^2/g}$
  (bottom).  The region below the straight magenta lines shows where
  inelastic processes may modify the picture based on elastic
  scattering for each type of system.}
  \label{fig:AMSB-pure-scattering-params}
\end{figure}

The thermally averaged transfer cross section, then, depends on four
parameters: $m_X$, $\Lambda$, $\alpha$, and $V_{\text{max}}$.  In
\figref{AMSB-pure-scattering-params}, we plot the ratio $\langle
\sigma_T \rangle /m_X$ in the $(m_X, \Lambda)$ plane for $\alpha = 1$
and three representative characteristic velocities: $V_\textrm{max} =
40~\units{km/s}$ for dwarfs, $V_\textrm{max} = 100~\units{km/s}$ for
LSBs, and $V_\textrm{max} = 1000~\units{km/s}$ for clusters.  For
masses $m_X \sim 1~\units{TeV}$ and $\Lambda\sim 10~\units{MeV}$, we
achieve transfer cross sections around the targeted range between
$0.1~\units{cm^2/g}$ and $1.0~\units{cm^2/g}$ for all three systems
under consideration.  The transfer cross section decreases as a
function of $v$ in the classical regime; thus, systems with larger
characteristic velocities have smaller cross sections, all else being
equal.  The LSB line at $0.1~\units{cm^2/g}$, for instance, lies below
that for dwarfs, because a larger interaction range (corresponding to a smaller
$\Lambda$) is needed to counter its larger velocity to give the same
$\sigma_T$ as the dwarfs. Toward the lower values of $m_X$, the
scattering exhibits resonant behavior due to the formation of
quasibound states~\cite{Tulin:2013teo}, analogous to Sommerfeld
enhancements in annihilations.

The region below the straight magenta lines in
\figref{AMSB-pure-scattering-params} is where the dark matter
typically has $(1/2) m_X v^2 > \Lambda$, and modifications from
inelastic scattering processes can be important.  We urge the reader
to keep in mind that while in this region the classical elastic
scattering cross section (for our assumed Yukawa potential) falls
below about $3\pi/\Lambda^2$, we expect other energy-exchange
mechanisms to become important in dark matter halos.  Note that for
clusters ($v \sim 3 \times 10^{-3}$), this is a substantial region of
the interesting parameter space: $(m_X/{\rm TeV}) \gtrsim
(\Lambda/ 10~{\rm MeV})$.  This suggests that the elastic glueballino
scattering curves plotted for clusters in
\figref{AMSB-pure-scattering-params} and other figures are far from
the whole story.  We expect new astrophysical phenomenology,
especially in clusters of galaxies, and this deserves separate
consideration.

\section{Glueballino relic density}
\label{sec:glueballinorelic}

One goal of supersymmetrizing the pure gauge hidden sectors considered
in \secref{glueball} is to revive the possibility of dark matter with
naturally the right relic density, as in the case of WIMPs, but now
for self-interacting dark matter.  In this section, we first review
the machinery required to calculate a glueballino relic density from
the freeze-out of thermal relic gluinos.  We then discuss the
possibility of realizing the correct thermal relic density through the
WIMPless miracle in AMSB models~\cite{Feng:2011ik}.

\subsection{Gluino freeze-out}
\label{sec:freezeout}

In a supersymmetric pure gauge hidden sector, the gluinos are
initially in equilibrium with a thermal bath of gluons at a hidden-sector
temperature $T^h$. As the Universe cools below the gluino mass
$m_X$, however, the gluinos freeze out.  The gluino is the lightest
supersymmetric particle in the hidden sector, and we will assume it is
stable. In the absence of couplings to the visible sector, stability
is guaranteed by Lorentz symmetry, as the gluino is the only fermion
in the hidden sector.

The gluino relic density is determined by the usual thermal freeze-out
analysis, but with the slight extra complication that it occurs in a
hidden sector with a temperature that may differ from the visible
sector.  For $S$-wave annihilation, the relic density of a thermal
relic in a hidden sector is~\cite{Feng:2008mu}
\begin{equation}
  \Omega_X \approx \frac{s_0}{\rho_{c0}}
  \frac{3.79 x_f}{(g_{*S}/\sqrt{g_*^\textrm{tot}}) \mplanck
\langle \sigma v \rangle } \ ,
\label{glueballinoRelic}
\end{equation}
where $s_0$ is the entropy of the visible sector today, $x_f \equiv
m_X/T_f$, $\rho_{c0}$ is the critical density today, and
$g_*^\textrm{tot}=g_* + \xi^4_f g_*^h$ at freeze-out.

We now discuss the various quantities entering
\eqref{glueballinoRelic}.  For the annihilation process $\tilde{g}
\tilde{g} \to g g$, we use an $S$-wave cross section,
\begin{equation}
  \langle \sigma v \rangle  = k_N \frac{\pi\alpha_X^2}{m_X^2} \ ,
\end{equation}
where $\alpha_X = g_h^2(m_X)/4\pi$ is the fine-structure constant with
a corresponding hidden-gauge coupling $g_h$ evaluated at the scale
$m_X$, and $k_N$ is an $\order{1}$ $N$-dependent coefficient, which we
simply set to $1$.  Additionally, we set $x_f=25\xi_f$, which is a
good approximation for a large set of parameters~\cite{Feng:2008mu}.
Given this, $\Omega_X$ scales approximately linearly with $\xi_f$.
The latest Planck results give a value of $\Omega_\textrm{DM} h^2 =
0.1196 \pm 0.0031$ from a six-parameter fit to the $\Lambda$CDM
model~\cite{Ade:2013zuv}.

To determine the number of relativistic degrees of freedom in the
visible and hidden sectors, note that, although the hidden and visible
sectors need to interact gravitationally, they do not necessarily have
to communicate otherwise, even at high energies.  Thus, the sector
temperatures are generically unrelated, and the ratio $\xi = T^h/T$
parametrizes this difference.  The comoving entropies in the visible
and hidden sector are independently conserved, and the values of $\xi$
at the different times $t_i$ and $t_f$ are related by
\begin{equation}
  \frac{g_{*S}^h(t_i)}{g_{*S}(t_i)} \xi_i^3
  = \frac{g_{*S}^h(t_f)}{g_{*S}(t_f)} \xi_f^3 \ .
  \label{eq:xiRelation}
\end{equation}
The effective numbers of relativistic degrees of freedom associated
with the entropy (energy) density in the visible and hidden sectors
are $g_{*S}$ and $g_{*S}^h$ ($g_*$ and $g_*^h$), respectively.  As we will see, for
most of the parameter space of interest, the gluinos freeze out at
visible-sector temperatures at or above $T_\textrm{SM}\approx
300~\units{GeV}$, so that all SM particles are relativistic and
$g_{*S}=g_*=106.75$.  Although there may be minimal supersymmetric
standard model (MSSM) superpartners with low enough masses to
contribute to $g_*$ at freeze-out, we assume the contribution is
negligible, with most of the visible supersymmetric-partner spectrum
being above $m_X$.  For the gluons and gluinos,
\begin{equation}
  g_*^h = g_{*S}^h =
  \begin{cases}
    4(N^2-1) & T^h \gtrsim m_X \\
    2(N^2-1) & m_X \gtrsim T^h > \Lambda \ .
  \end{cases}
\end{equation}

\subsection{The WIMPless miracle and AMSB}
\label{sec:wimpless}

As noted above, the gluino thermal relic density has the parametric
dependence
\begin{equation}
  \Omega_X \sim \frac{1}{\langle \sigma v \rangle }
\sim \frac{m_X^2}{\alpha_X^2} \ .
\end{equation}
For weak-scale masses and weak interaction coupling strengths,
$\Omega_X$ is of the desired size; this is the essence of the WIMP
miracle.  For the hidden sector, we have great freedom in choosing the
parameters $m_X$ and $\alpha_X$, and may simply choose them to yield
the correct relic density.  However, it is preferable if the correct
mass-to-coupling ratio is set in a noncontrived way.  This is a
property of models that realize the WIMPless
miracle~\cite{Feng:2008ya,Feng:2008mu}, where the dark matter mass
and coupling are not fixed individually, but the ratio
$m_X^2/\alpha_X^2$ is fixed to the desired value by the model
framework.

Supersymmetric models with AMSB~\cite{Giudice:1998xp,Randall:1998uk}
provide a particularly nice realization of the WIMPless
miracle~\cite{Feng:2011ik,Feng:2011uf,Feng:2011in}.  In AMSB, the MSSM
is sequestered from the supersymmetry-breaking sector, so the gaugino
masses in the visible sector do not receive any tree-level
contributions and are instead generated by the Weyl anomaly, leading
to
\begin{equation}
  m_v \sim \frac{\alpha_v}{4\pi} m_{3/2} \ ,
\end{equation}
where $m_{3/2}$ is the gravitino mass, $\alpha_v$ is a SM
fine-structure constant, and $m_v$ is of the order of the weak scale,
if these models are to address the gauge hierarchy problem.  In any
additional sequestered hidden sector of the theory, the hidden-sector
superpartner masses will be given by a similar relation,
\begin{equation}
  m_X \sim \frac{\alpha_X}{4\pi} m_{3/2} \ ,
\end{equation}
where $\alpha_X$ is the hidden sector's fine-structure constant.
Since there is only one gravitino mass, $m_X / \alpha_X \sim m_v /
\alpha_v$, and any hidden sector thermal relic in AMSB can be expected
to have the desired relic density, even if $m_X$ and $\alpha_X$
differ, perhaps greatly, from the SM values.

The visible sector of AMSB models contains a stable thermal relic, the
lightest neutralino.  However, the standard AMSB relations imply that
this is the wino.  Winos annihilate very efficiently, and must have
masses around $2.7-3.0~\tev$ to be all of dark
matter~\cite{Hisano:2006nn,Hryczuk:2010zi}.  The thermal relic density
scales as $\sim m_{\tilde{W}}^{-2}$, and so for lighter and more
natural values closer to the Large Electron-Positron (LEP2) collider
experimental limit $m_{\tilde{W}}
\agt 100~\gev$~\cite{lep2-01-03.1,lep2-02-04.1}, the wino thermal
relic density is completely negligible.  We will therefore neglect it
below, and take this as additional motivation to develop AMSB models
with viable hidden-sector dark matter candidates.

The particle spectrum in AMSB models is completely specified by
quantum numbers, dimensionless couplings, and the gravitino mass.  In
the visible sector, the wino mass limit $m_{\tilde{W}} \gtrsim
100~\units{GeV}$ implies
\begin{equation}
  m_{3/2} \gtrsim 37~\units{TeV} \ .
  \label{eq:LEP2-bound}
\end{equation}
In the hidden sector, at scales above $m_X$, the one-loop
$\beta$-function coefficient is $b_H=-3N$, and the theory is
asymptotically free.  The gluino mass is
\begin{equation}
  m_X = -b_H \frac{\alpha_X}{4\pi} m_{3/2} =
3N \frac{\alpha_X}{4\pi} m_{3/2} \ .
\end{equation}
Below $m_X$, we have a nonsupersymmetric SU($N$) gauge theory with a
$\beta$-function coefficient: $b_L=-(11/3)N$.  The theory is expected
to confine at the scale
\begin{equation}
  \Lambda \sim m_X \exp\left(\frac{-6\pi}{11 N\alpha_X}\right)
  = m_X \exp\left(\frac{-9m_{3/2}}{22 m_X}\right) \ .
\label{Lambda}
\end{equation}
With this relationship, the relic density in~\eqref{glueballinoRelic}
becomes
\begin{equation}
  \Omega_X \approx \frac{s_0}{\rho_{c0}}
  \frac{\left[g_* + 2(N^2-1)\xi_f^4\right]^{1/2}}{g_{*S}}
  \frac{3.79 \cdot 25 \xi_f}{\mplanck}
  \frac{9 N^2}{16 \pi^3} m_{3/2}^2  \ .
\label{glueballinoRelic-reduced}
\end{equation}

\section{Glueballino/Glueball dark matter without connectors}
\label{sec:noconnectors}

Given the results above, we can now present simple AMSB models of
self-interacting dark matter.  We begin by considering the simple case
without connector fields, in which the visible and hidden sectors are
decoupled.  The visible sector is the MSSM; the tachyonic slepton
problem is assumed to be solved in a way that does not impact the
masses of the MSSM gauginos, and the wino is assumed to be the visible
lightest supersymmetric particle (LSP), with a negligible thermal relic
density.  The hidden sector is a
pure SU($N$) gauge theory, consisting of gluinos and gluons, which
confine to form glueballino and glueball dark matter.

There are only four independent parameters in the theory, which may be
taken to be
\begin{equation}
m_X, ~\Lambda, ~N, ~\xi_f \ .
\end{equation}
These determine $\alpha_X$ and $m_{3/2}$ through \eqref{Lambda}. In
contrast to the model-independent discussion of
\secref{glueballinoscattering}, in AMSB models, renormalization group
equations relate the high-scale parameters $m_X$ and $\alpha_X$ to the
low-scale parameter $\Lambda$.  In terms of these parameters, the
glueball self-interaction cross section and relic density are
determined as described in
\secsref{glueballscattering}{glueballrelic}, and the glueballino
self-interaction cross section and relic density are determined as
described in \secsref{glueballinoscattering}{glueballinorelic}.

\begin{figure}[t]
  \centering
  \includegraphics[width = 0.69 \textwidth]{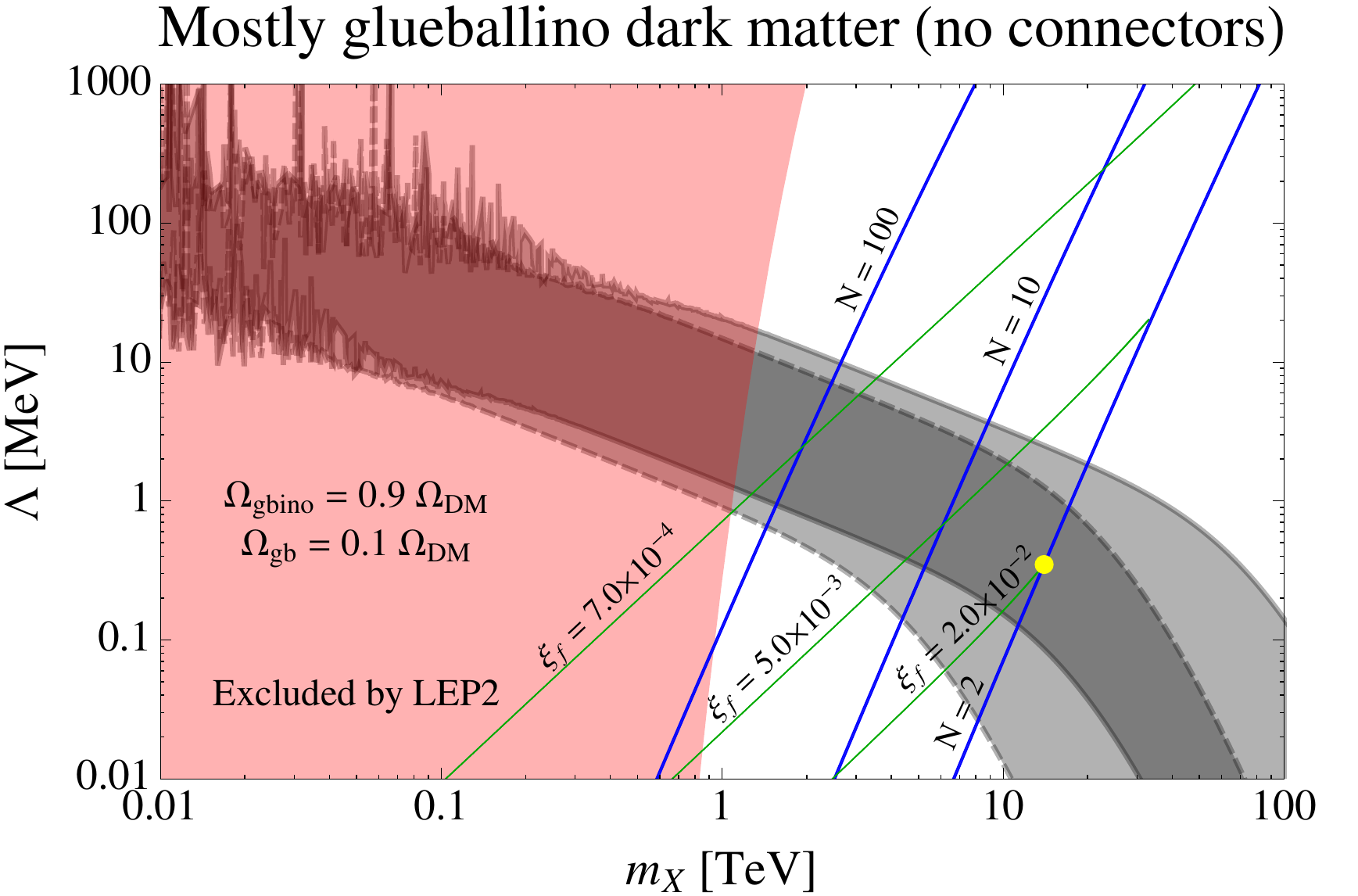} \\
\vspace*{0.15in}
    \includegraphics[width = 0.69 \textwidth]{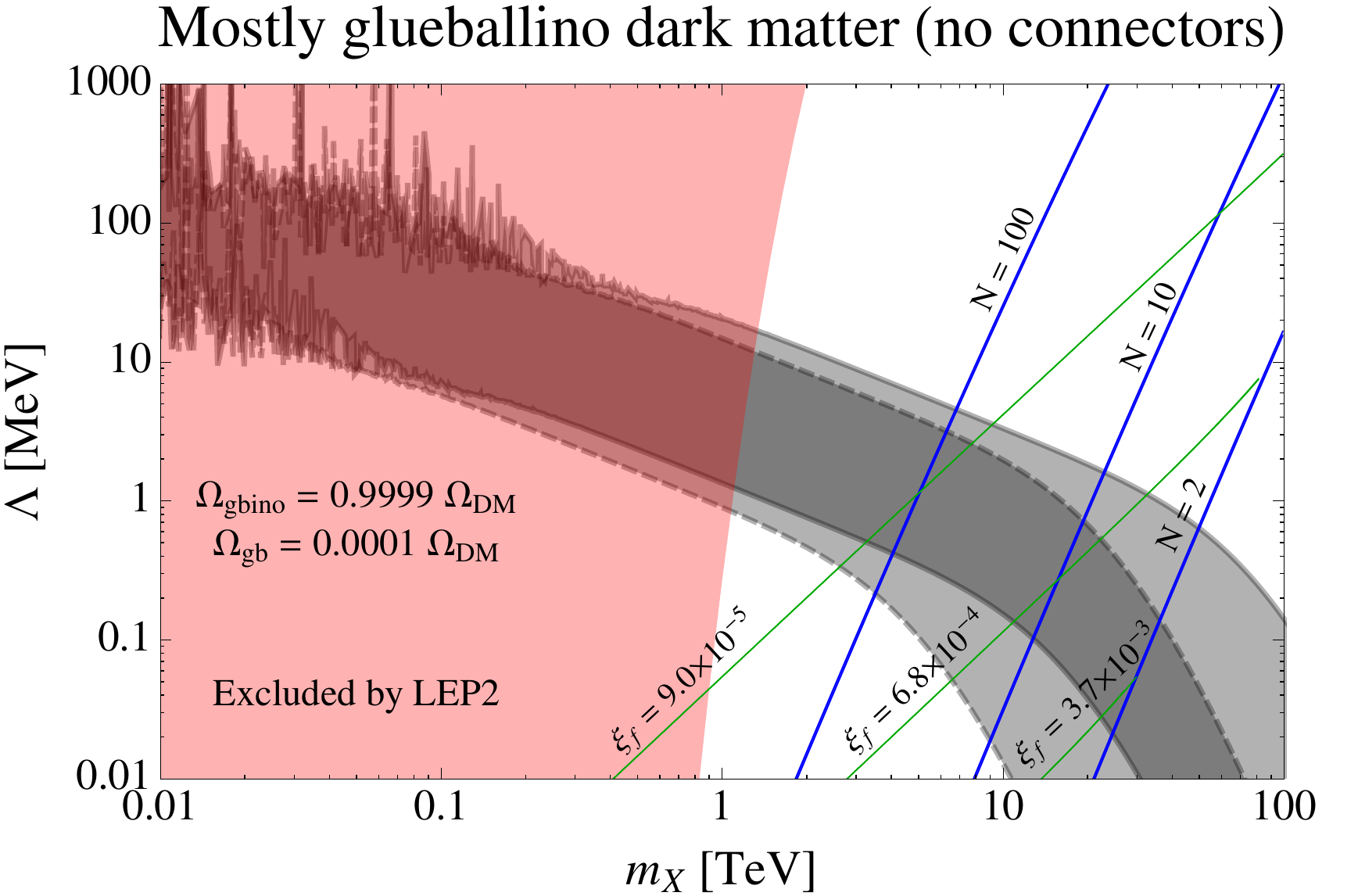}
  \caption{Mostly glueballino dark matter in AMSB models with
      pure SU($N$) hidden sectors without connectors.  Glueballinos
    make up 90\% (top) or 99.99\% (bottom) of the dark matter, and
    glueballs make up the remaining portion.  For a point in the
    $(m_X, \Lambda)$ plane, these constraints on the relic densities
    determine $N$ and $\xi_f$; contours of constant $N$ and $\xi_f$
    are shown.  The gray shaded bands are from
    \figref{AMSB-pure-scattering-params} and give the regions where
    the glueballino self-interaction cross section is in the preferred
    range. The red shaded region is excluded by null searches for
    visible-sector winos at LEP2.  The yellow dot in the top panel
    defines a representative model with $m_X \simeq 14~\units{TeV}$,
    $\Lambda \simeq 0.35~\units{MeV}$, $N=2$, and $\xi_f \simeq 0.02$.
  }
  \label{fig:AMSB-pure-relic}
\end{figure}

We first present results for models with mostly glueballino dark
matter in \figref{AMSB-pure-relic}.  We scan over the $(m_X, \Lambda)$
plane.  At every point in this plane, we require that glueballinos
make up 90\% (top panel) or 99.99\% (bottom panel) of the dark matter,
and glueballs make up the remaining 10\% or 0.01\%.  The constraints
on $\Omega_{\gbino}$ and $\Omega_{\gb}$ determine $N$ and $\xi_f$;
contours of constant $N$ and $\xi_f$ are shown.  The lower bound of
\eqref{eq:LEP2-bound} excludes parameter space with a low $m_X$.  In the
remaining parameter space, $m_X/\Lambda \gtrsim 10^3$, which is more than
sufficient to ensure $T_f^h > \Lambda$, so gluino freeze-out occurs in
the weakly interacting theory, and the thermal freeze-out calculation
is valid.

These relic density results for a particular glueballino density may
be understood as follows.  On a given curve of constant $N$, larger
dark matter masses imply larger thermal relic densities and so require
smaller values of $\xi_f$ to keep $\Omega_{\gbino}$ fixed. Once
$\xi_f$ decreases, a larger $\Lambda$ is required to keep
$\Omega_{\gb}$ constant.  Note also that for $\Lambda \sim \mev$ and
$\xi_f \sim 1$, glueballs overclose the Universe.  To avoid this,
$\xi_f$ must be lowered, and, to have mostly glueballino dark matter,
$m_X$ must be a bit larger than the weak scale.  In short, the
presence of glueballs forces the model away from the {\em a priori}
most natural parameter space with a low $m_X$ and $\xi_f \sim 1$.  In
the context of AMSB, however, it is rather natural to assume that the
hidden and visible sectors are separated at high scales and $\xi_f \ll
1$.  Given this, the WIMPless miracle is nicely realized in regions of
parameter space with $\xi_f \sim 0.01$ and $N \sim \order{1}$ for
$\Omega_{\gbino} = 0.9 \Omega_\text{DM}$.

There are also differences between the 90\% and 99.99\% glueballino
cases.  The curves of constant $N$ and constant $\xi_f$ shift as the
relative amounts of glueball and glueballino dark matter change.  By
focusing on a particular point in the $(m_X,\Lambda)$ plane and
comparing \eqref{eq:gbRelic} and \eqref{glueballinoRelic-reduced},
we find
\begin{equation}
  \xi_f \sim \frac{\Omega_\gb^{1/2}}{\Omega_\gbino^{1/2}}
  \qquad\textrm{and}\qquad
  N \sim \frac{\Omega_\gbino^{3/4}}{\Omega_\gb^{1/4}}
\end{equation}
for $N^2 \gg 1$. When the glueball density is reduced by three orders of
magnitude, we expect $N$ to increase by a factor of $10^{3/4} \sim 6$
and $\xi_f$ to decrease by a factor of $10^{3/2} \sim 30$; this can be
seen in \figref{AMSB-pure-relic}.

Of course, the goal is not simply to obtain a multicomponent model of
dark matter with the correct relic densities, but to obtain
self-interacting dark matter.  The regions with the preferred
self-interaction cross sections are also shown in
\figref{AMSB-pure-relic}.  For values of $m_X \sim 10~\tev$, $\Lambda
\sim 1~\mev$, $2 \leq N \alt 10$, and $10^{-3} \alt \xi_f \alt
10^{-2}$, we find models that satisfy the relic density constraints
and also satisfy the scattering constraints for dwarfs and
LSBs. Viable models also exist for the lower values of $m_X$ down to the LEP2 limit
for larger values of $N$ and lower values of $\xi_f$. A representative model is one with
$m_X \simeq 14~\units{TeV}$, $\Lambda \simeq 0.35~\units{MeV}$, $N=2$,
and $\xi_f \simeq 0.02$; this is shown as a yellow dot in
\figref{AMSB-pure-relic}. For these parameters, \figref{timeline1}
shows how the dark matter coupling behaves from the scale $m_X$ down
to confinement.

Measurements of nuclei abundances and of the CMB place restrictions on
the number of light degrees of freedom $\Neff$ around the time of BBN
that contribute to the expansion of the Universe.  Results from Planck
give $\Neff=3.30 \pm 0.27$~\cite{Ade:2013zuv}.  An interesting
question, then, is whether these models also imply nonstandard values
of $\Neff$.  Once the hidden-sector temperature drops below the
confinement scale, glueballinos and glueballs form.  This occurs when
the visible sector's temperature is
\begin{equation}
  T_\Lambda = \frac{T_\Lambda^h}{\xi_\Lambda} \sim \frac{\Lambda}{\xi_\Lambda}
  = \frac{\Lambda}{\xi_f} \left(\frac{g_{*S}(t_f)}{g_{*S}(t_\Lambda)}\right)^{1/3} \ ,
  \label{eq:confinementTemp}
\end{equation}
using \eqref{eq:xiRelation} with $g_{*S}^h(t_\Lambda)=g_{*S}^h(t_f)$.
For the representative example parameters given above, the confinement
scale is $T_\Lambda \sim 90~\units{MeV}$; confinement occurs well
before BBN and structure formation, as expected.  There is therefore
no relativistic, massless species to act as the hidden-sector bath
during BBN.  At the time of BBN, the hidden-sector temperature is not
well defined, and its contribution to $\Neff$ is essentially zero.

\begin{figure}[t]
  \centering
  \includegraphics[width=0.69 \textwidth]{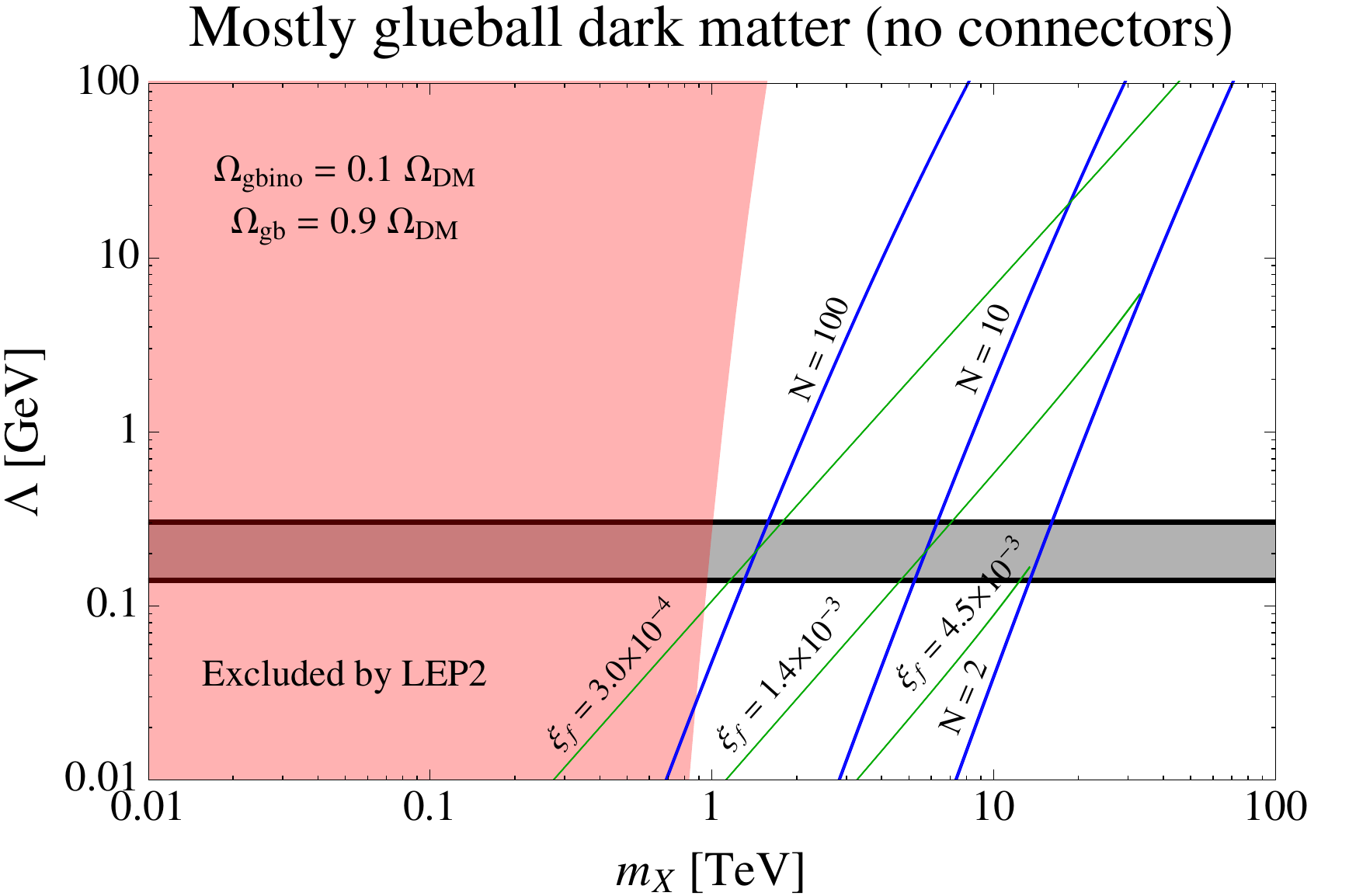} \\
\vspace*{.15in}
    \includegraphics[width=0.69 \textwidth]{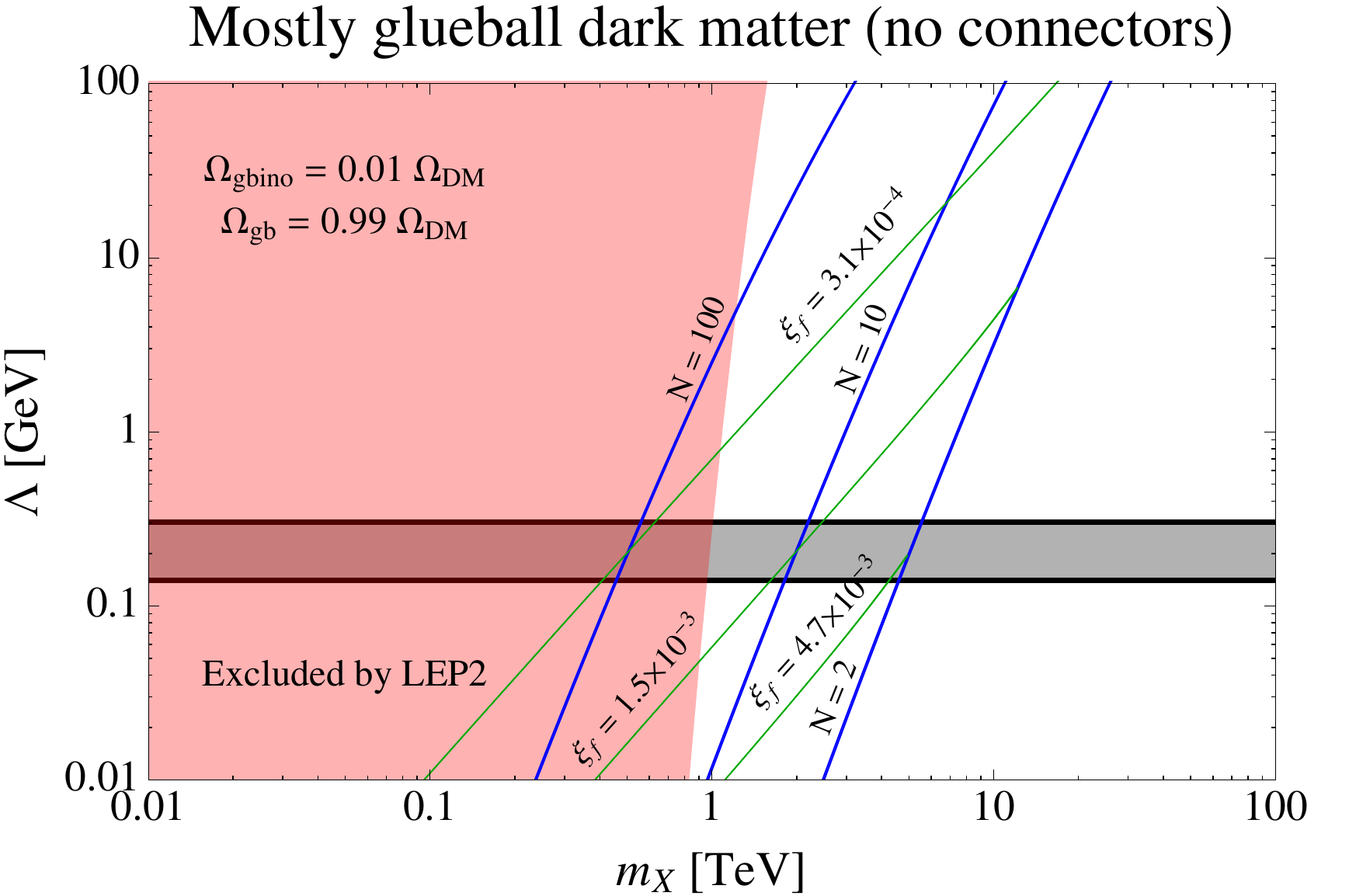}
  \caption{Mostly glueball dark matter in AMSB models with pure
      SU($N$) hidden sectors and no connectors.  Glueballs make up
    90\% (top) or 99\% (bottom) of the dark matter, and glueballinos
    make up the remaining portion. For a point in the $(m_X, \Lambda)$
    plane, these constraints on the relic densities determine $N$ and
    $\xi_f$; contours of constant $N$ and $\xi_f$ are shown.  The
    gray shaded band is from \figref{glueball} and gives the region
    where the glueball self-interaction cross section is in the
    preferred range.  The red shaded region is excluded by null
    searches for visible-sector winos at LEP2.}
  \label{fig:AMSB-pure-relic-glueball}
\end{figure}

We next consider the case of mostly glueball dark matter.  To be
concrete, we present the case of $\Omega_{\gb} = 0.9 \,
\Omega_{\text{DM}}$ and $\Omega_{\gbino} = 0.1 \, \Omega_{\text{DM}}$
in \figref{AMSB-pure-relic-glueball}.  Once again, we show contours of
constant $N$ and $\xi_f$, but now we include the glueball scattering
constraints from \figref{glueball}, since glueballs are the dominant
component of dark matter.  The values of $m_X$ that satisfy relic and
scattering constraints for a given $N$ are fairly similar to those in
the case of mostly glueballino dark matter; however, the corresponding
values of $\Lambda$ are a few orders of magnitude larger than the
mostly glueballino case.

\begin{figure}[t]
  \centering
  \includegraphics[scale=0.7]{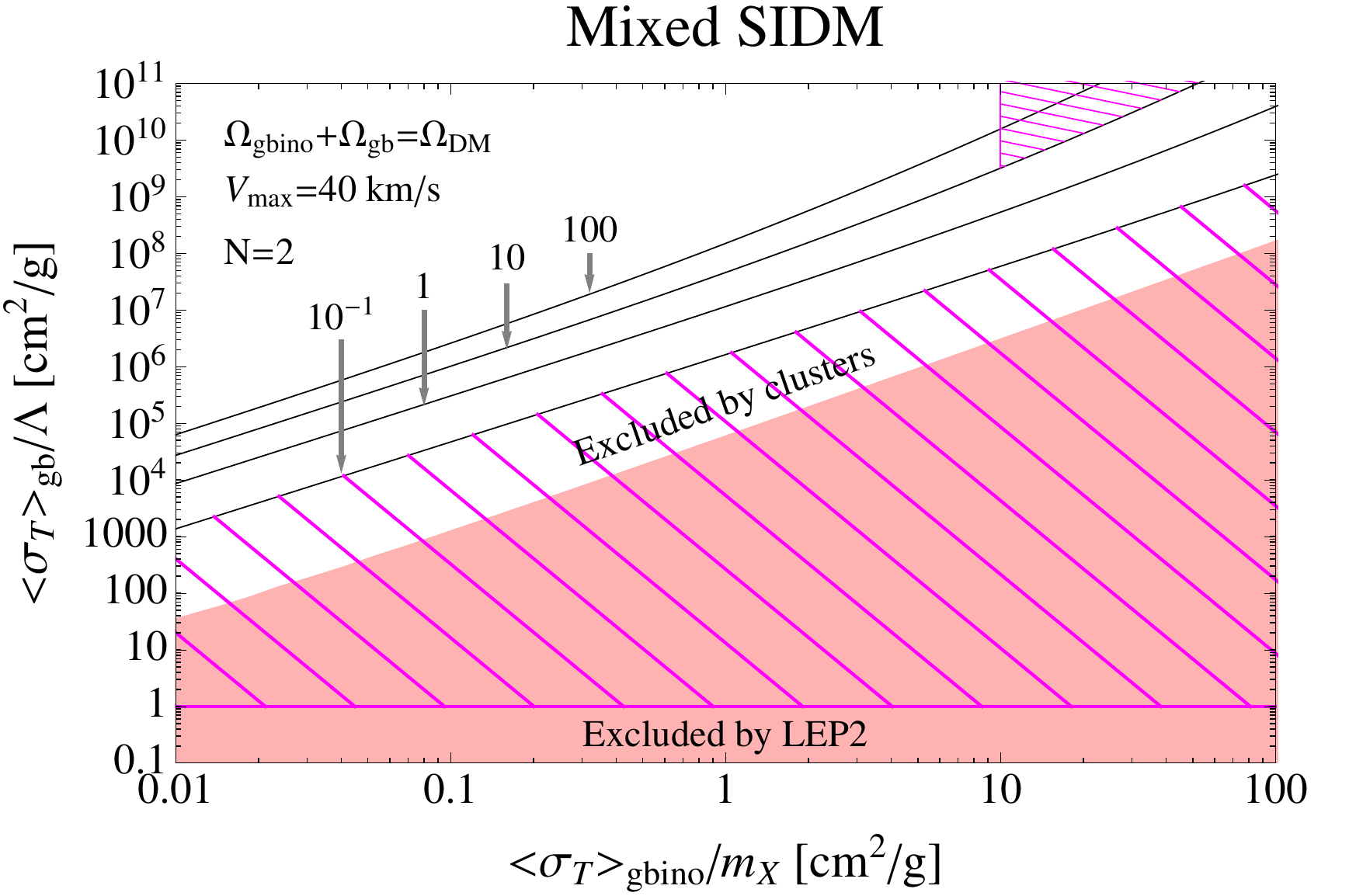}
  \caption{Mixed dark matter without connectors to the standard
      model.  We show curves of constant
    $\Omega_{\gbino}/\Omega_{\gb}$ in the $(\langle \sigma_T
    \rangle_{\gbino} /m_X, \langle \sigma_T \rangle_{\gb} /\Lambda)$
    plane, for $N=2$ and considering dwarf systems with
    $V_\textrm{max}=40~\units{km/s}$. The black curves have
    $\Omega_{\gbino}/ \Omega_{\gb} = 0.1, 1, 10, 100$, as indicated.
    The bound from LEP2 is shown in the red shaded region.  A
    stronger bound from clusters is shown in the lower hatched
    magenta region; since the glueball scattering cross section is the
    same on all scales, its value is limited for the dwarf systems to
    avoid violating bounds from cluster scales. We caution the reader
    that the bound may be stronger and it is certainly not as sharp as
    indicated by the hatched region. The hatched magenta wedge near
    the upper right-hand portion of the graph represents an upper
    limit of $10~\units{cm^2/g}$ for the case of mostly glueballino
    dark matter, which will have important implications for cores in
    dwarfs galaxies and may be excluded by a comparison to the
    observed core sizes and densities (e.g.,
    Ref.~\cite{Rocha:2012jg,zavala13}). }
  \label{fig:AMSB-pure-relic-Omega}
\end{figure}

In \figsref{AMSB-pure-relic}{AMSB-pure-relic-glueball}, the fraction
of glueballino to glueball dark matter is fixed.  Of course, different
values are possible.  In \figref{AMSB-pure-relic-Omega}, we fix $N=2$
and vary $m_X$ and $\Lambda$; $\xi_f$ is set by the requirement that
$\Omega_{\gbino} + \Omega_{\gb} = \Omega_{\text{DM}}$.  The results
are presented in the $(\langle \sigma_T \rangle_{\gbino} / m_X,
\langle \sigma_T \rangle_{\gb} / \Lambda)$ plane, where
$V_{\text{max}} = 40~\km/\s$, and contours of constant
$\Omega_{\gbino} / \Omega_{\gb}$ are shown.  Regions excluded by LEP2
and by cluster bounds are shaded.

\Figref{AMSB-pure-relic-Omega} shows that the fraction of dark matter
that is composed of glueballinos may take almost any value in the parameter space.
Of course, regions of parameter space that are overwhelmingly
glueballino dominated and have too-large glueballino self interactions
are excluded, as are regions that are overwhelmingly
glueball dominated with too-large glueball self interactions.  The
parts of parameter space that are certainly excluded by these
considerations are indicated, but the position of this boundary is
somewhat uncertain and requires detailed $N$-body simulations
(modeling both components of dark matter) to determine.  The cluster
constraints~\cite{Randall:2007ph,Peter:2012jh} are relevant here
because glueballs have a velocity-independent scattering cross section
and these constraints dictate that glueballs must be the subdominant
component of dark matter in all of the parameter space shown in
\figref{AMSB-pure-relic-Omega}.

Especially interesting, however, are the regions of parameter space
with a subdominant component of dark matter that self interacts very
strongly.  For example, the dark matter may be 99\% glueballinos and
1\% glueballs, but the glueballs may have $\langle \sigma_T
\rangle_{\gb} / \Lambda \sim 10^5 - 10^{11}~\cm^2/\g$.  Such
possibilities are not ruled out by the constraints discussed so far
but may have very interesting astrophysical implications.

It has been pointed out that, at early times before the halo has had
time to form a core through self interactions, seed black holes can
grow by accreting self-interacting dark
matter~\cite{Hennawi:2001be}. In the mixed self-interacting dark
matter scenario where one of the components has $\langle
\sigma_T\rangle/m \gg 1\cm^2/\g$, this accretion can be highly
enhanced.  The possibility that supermassive black hole growth is
seeded by the self interactions of either the glueballs or
glueballinos is an exciting prospect. There is not yet a clear picture
of how $10^9~{\rm M}_\odot$ quasars are assembled already by $z
\gtrsim 6$ within the standard $\Lambda$CDM cosmology. Models starting
with the expected $100~{\rm M}_\odot$ seeds require special
assumptions about the mass accretion histories of these
quasars~\cite{oai:arXiv.org:0807.4702}, which become more strained as
higher redshift quasars are found~\cite{oai:arXiv.org:1106.6088}.
Self interactions within the dark matter sector may have a big role to
play in this story, as they generically enhance the early black hole
accretion rate.

There is a tight correlation between the mass of supermassive black
holes in the centers of galaxies and the velocity dispersion or
luminosity of the bulge~\cite{oai:arXiv.org:0903.4897}. By requiring
that the predicted masses of supermassive black holes are not overly
large, it should also be possible to constrain the ratio
$\Omega_\gb/\Omega_\gbino$ in mixed self-interacting dark matter
models where $\langle \sigma_T \rangle_\gb/\Lambda$ is large.  To
correctly implement this constraint, many new features of our simple
model and their astrophysical consequences will have to be worked
out. We highlight a few of these below.

The details of the capture of glueballs by a seed black hole will differ
significantly from the treatment in Ref.~\cite{Hennawi:2001be}. The
black hole capture depends sensitively on the density profile of
glueballs, and this is tightly correlated with the potential well of
the galaxy, which is dominated by glueballinos. In particular,
although an isolated strongly self-interacting dark matter halo will
undergo core collapse, this is not true when the strongly
self-interacting component (glueballs) is a small fraction of the dark
matter.

A complicating factor is that the glueballs and glueballinos will
scatter off of each other. Each collision will change the velocity of
glueballs by ${\cal O}(1)$, but the velocity of glueballinos will only
change by $\Lambda/m_X \ll 1$. The glueballino-glueball scattering
cross section should be of the order of the geometric cross section
($\sim 1/\Lambda^2$), and, thus, this could be an important effect if
the number density of glueballinos is much larger than that of
glueballs (either because of a small $\Omega_\gb/\Omega_\gbino$ or as
glueballs are depleted due to accretion by the black
hole). Conversely, this scattering could also have an impact on the
glueballino density profile if the number density of glueballs is
large enough to overcome the small momentum transfer.

Another important effect, relevant for halo properties as well as
black hole growth, is cooling. We have focused on elastic collisions in
this paper, but as mentioned previously there are also inelastic
processes leading to cooling through the emission of
glueballs. Cooling will funnel more glueballs into the inner regions
(modulo angular momentum constraints) and increase the black hole
accretion rate. Note that, unlike the baryons, competing effects from
star formation and subsequent heating by UV photons are not relevant
for glueballs.

As an extreme example, one could assume that all of the glueballs are
bound up in the central supermassive black hole. In this case, we can
use measured ratios of the black hole masses to halo masses to put an
upper limit on $\Omega_\gb/\Omega_\gbino$.  For the Milky Way, this
ratio is $\sim 10^{-5}$, while for Andromeda the ratio is more like
$10^{-4}$.  (It should be kept in mind that the black hole will also
accrete baryons and grow, so this is a lenient upper bound.)  Rather
than focus on the Local Group, one could look more generally at the
black hole mass--virial mass relation for all galaxies, but as
expected there is a lot of scatter in this
relation~\cite{2012MNRAS.419.2497B}.

To illustrate the effect of these constraints on the model parameter
space, we have shown two possibilities in \figref{AMSB-pure-relic}:
one with $\Omega_\gb/\Omega_\gbino = 0.1$ (which may not be viable
given the arguments above) and a second with $\Omega_\gb/\Omega_\gbino
= 10^{-4}$. There is no impediment in making this ratio even smaller,
although there is no natural reason to do so. In addition, as
$\Omega_\gb/\Omega_\gbino$ is reduced, the regions with small $N$ move
into the regime where the inelastic process will be important for all
relevant velocities (dwarfs to clusters).

\section{Glueballino/Glueball dark matter with connectors}
\label{sec:connectors}

Although a pure SU($N$) hidden sector with no connectors can
accommodate both early Universe and structure formation constraints,
it is interesting to consider the possibility of connector fields that
allow communication between the hidden and visible sectors.  Such
scenarios may have, of course, a larger number of testable
implications.  In addition, as we will see, if the connectors mediate
annihilation or decays to the visible sector, the viable parameter
space may be significantly altered.

\begin{figure}[t]
  \centering
  \includegraphics[scale=0.7]{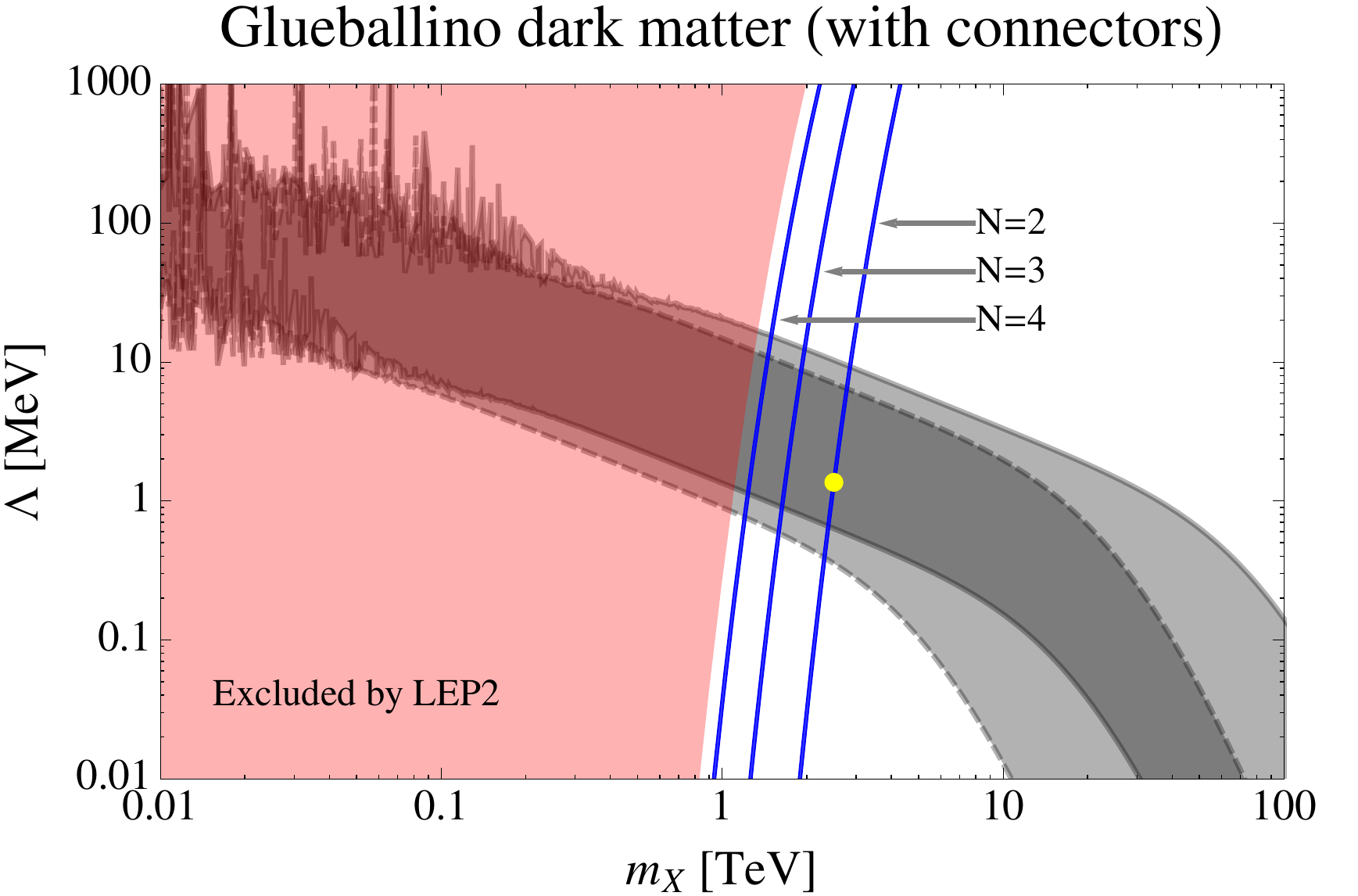}
  \caption{Glueballino dark matter in AMSB models with pure
      SU($N$) hidden sectors and connectors to the SM.  Glueballinos
    are assumed to make up all of the dark matter. The relic density
    constraints are given in the $(m_X,\Lambda)$ plane with $\xi_f=1$;
    contours of constant $N$ are shown.  The gray shaded bands are
    from \figref{AMSB-pure-scattering-params} and give the regions
    where the glueballino self-interaction cross sections are in the
    preferred range. The red shaded region is excluded by null
    searches for visible-sector winos at LEP2.  The yellow dot defines
    a representative model with $m_X \simeq 2.5~\units{TeV}$, $\Lambda
    \simeq 1.4~\units{MeV}$, and $N=2$.  }
  \label{fig:AMSB-pureConn-relic}
\end{figure}

If the hidden and visible sectors communicate, we expect the
temperatures of the two sectors to coincide nearly until kinetic
decoupling at confinement.  If glueballs are stable, they will
generically overclose the Universe, and so there must be a mechanism
to reduce the glueball density.  Let us assume that this mechanism
exists and reduces the glueball relic density to a negligible level.
We can then immediately determine the consequences for the parameter
space.  For a given point in the $(m_X, \Lambda)$ plane with $\xi_f =
1$, there are contours of constant $N$ on which $\Omega_{\gbino} =
\Omega_{\text{DM}}$. These are shown in \figref{AMSB-pureConn-relic},
along with the self-interaction constraints.  We see that the LEP2
bound excludes all but the $N \leq 4$ possibilities, but now, for small
$N$, the allowed values of $m_X$ are much reduced and more natural
relative to the case without connectors.

A straightforward way to eliminate glueballs is through decays, but
other constraints render this scenario unacceptable. The glueballs
have a mass around $1$ to $10~\units{MeV}$, so possible decay products
will be photons, electrons, and neutrinos.
Decays to photons will typically take too long and happen well after
BBN. If too much energy and entropy is injected into the visible
sector at $T \lesssim 1~\units{MeV}$, then there is an unacceptably
large contribution to $\Neff$.  Decays to electrons after
$1~\units{MeV}$ face a similar problem, and, in addition, they can
break up deuterium and ruin its BBN abundance (if the glueball is
heavy enough).  Decays to light neutrinos are problematic because
glueballs can be produced in supernovae, escape the neutrino sphere,
and cool the supernovae too efficiently.  If we attempt to adjust
parameters to get around the difficulties with either electrons or
neutrinos, then we encounter problems with $e^+ e^-$ collider
constraints.  We are led to consider alternative processes to
eliminate the glueball density.

Since decays after confinement are highly constrained, we investigate
reducing the glueballino density by depleting the gluon density before
confinement. The gluons may annihilate to SM particles via loop
diagrams, but the reverse process needs to be suppressed. Let us
introduce a right-handed neutrino $\nu_R$. The $\nu_R$ is a SM gauge
singlet with a mass, $m_R\sim\units{GeV}$, and it could be one of the sterile
states in a seesaw mechanism to produce neutrino masses.  Our goal is for
the gluons to annihilate into right-handed neutrinos, which then decay
quickly into SM particles before they can annihilate back into gluons.

To implement this scenario, we postulate that there is a connector
field, $C$, with a mass $m_C$ that allows communication between the hidden
and visible sector. The connector has a Yukawa interaction, $\lambda_R
C \bar{\nu}_R \nu_R$, in the visible sector and a gauge interaction
with the gluons with a strength $g_h$ in the hidden sector.  Integrating
out the connector produces the effective interaction
\begin{equation}
  \mathcal{L} \sim \frac{1}{16\pi^2} \frac{\lambda_R^2 g_h^2}{m_C^3}
  G^h_{\mu \nu} G^{h\, \mu \nu} \bar{\nu}_R \nu_R \ .
\end{equation}
This interaction leads to an annihilation cross section,
\begin{equation}
  \langle \sigma v \rangle_{g g \to \bar{\nu}_R \nu_R} \sim
  \frac{\lambda_R^4 g_h^4}{8\pi (16\pi^2)}
  \frac{T^4}{m_C^6} \equiv \sigma_0 z^{-4} \ ,
\end{equation}
where $z=m_R/T$.  Note that the annihilation of gluons into
right-handed neutrinos is subdominant to the annihilation rate of
gluons into gluinos and can be ignored in the gluino freeze-out
calculations. The right-handed neutrino decays with a rate of
\begin{equation}
  \Gamma_R \sim \frac{g_\nu^2}{4\pi} \frac{m_R^2}{T} \equiv \Gamma_0 z
\end{equation}
into SM particles at the tree level with a coupling strength $g_\nu$.  As
long as the neutrino decay rate is much faster than the gluon
annihilation into neutrinos (and both are faster than the Hubble
expansion), the gluons cannot maintain their equilibrium density, and
their energy is transferred to SM particles.  The depletion terminates
no later than $\sim m_R$, when any surviving right-handed neutrinos
freeze out, and the gluon density decreases subsequently only due to
Hubble expansion.

To give a concrete example, consider the following parameters: $N=2$,
$m_X=2.5~\units{TeV}$, $\Lambda\simeq 1.4~\units{MeV}$, $m_C =
0.5~\units{TeV}$, $m_R = 1~\units{GeV}$, $g_h=1.1$, $\lambda_R=1.6$,
and $g_\nu=0.1$. The output glueball relic density is $\sim$ 5\% of
the total dark matter abundance.  We find this result by numerically
solving the coupled Boltzmann equations for the gluons and
right-handed neutrinos:
\begin{eqnarray}
  Y_g'(z) &=& -z^{-6} \sigma_0 \frac{s(m_R)}{H(m_R)}
  \left(Y_g^2 - Y_R^2 \right) \\
  Y_R'(z) &=& -z^{-6} \sigma_0 \frac{s(m_R)}{H(m_R)}
  \left(Y_R^2 - Y_g^2 \right)
  - z^2 \frac{\Gamma_0}{H(m_R)} Y_R \ ,
\end{eqnarray}
where $s(m_R)$ and $H(m_R)$ are the entropy and Hubble rate at
$T=m_R$.  The initial conditions $Y_R(z_f)$ and $Y_g(z_f)$ are given
by \eqref{eq:Yinf} at the dark matter freeze-out: $z_f=25 m_R/m_X$.  These
differential equations tend to be fairly stiff, so, in certain regions
of parameter space, it is beneficial to decouple the equations. We may
do so if the neutrino decay term dominates, allowing us to approximate
$Y_R$ as exponentially decaying.  Solving the decoupled differential
equation yields results that are numerically similar (typically within
10\%) to solving the full set of coupled equations when the decay term
dominates.

There are few constraints on this mechanism.  Prior to confinement, a
large amount of entropy is transferred from the gluons to light SM
particles.  Since the right-handed neutrinos are still relativistic,
there is no entropy nonthermally deposited into the visible sector.
All the right-handed neutrino decay products fall into equilibrium
with the bath well before BBN.  With a nonzero glueball density, a
concern might be that the glueballs will be able to decay to SM
particles via off-shell right-handed neutrinos and nonthermally
deposit entropy into the visible sector.  If the right-handed neutrino
decays into a left-handed neutrino and the Higgs, then we expect the
glueball decay rate into $\bar{\nu}_L \nu_L e^+ e^- e^+ e^-$ to be
\begin{equation}
  \Gamma_\textrm{gb} \sim y_e^4 g_\nu^4 \frac{\Lambda^{19}}{m_C^6 m_h^8 m_R^4} \ ,
\end{equation}
where $y_e$ is the electron Yukawa coupling and $m_h$ is the mass of
the Higgs.  This decay rate is slow enough that the glueballs are
essentially stable and, further, will not contribute significantly to
$\Neff$, since they are nonrelativistic below $1~\units{MeV}$, and
they will not have a large impact on the expansion rate of the
Universe during BBN, given their small energy density.  Our glueball
depletion process is robust, and it is consistent with terrestrial and
cosmic constraints \cite{Abazajian:2012ys}.

\section{Conclusions}
\label{sec:conclusions}

We have explored the possibility that dark matter may be a composite
particle, made up of bound states of a dark analogue of QCD in the
hidden sector.  Such constructions lead to rich and varied phenomena
that are distinct from the WIMP scenario more typically considered.
It also naturally leads to large self interactions of the dark matter,
which can explain several observational puzzles in the small-scale
structure of the Universe.

The simplest scenarios contain only dark gluons, which confine into
glueballs with cosmologically interesting scattering cross sections
for confinement scales around 100 MeV.  Arranging the correct relic
density requires one to disconnect the temperatures in the hidden and
visible sectors such that their ratio at confinement is $\sim
10^{-3}$.

A richer theory arises when one considers supersymmetric versions, for
which the dark gluino mass provides a separate mass scale and (in
AMSB) can provide the correct relic density of glueballinos via the
WIMPless miracle.  The phenomenology depends crucially on how
connected the hidden sector is to the visible matter.  If there are no
light connecting particles, one can dial the balance of dark matter
from glueballs to glueballinos by adjusting the relative temperatures
of the hidden and visible sectors.  These mixed scenarios are
strongly interacting analogues of atomic dark
matter~\cite{Kaplan:2009de,CyrRacine:2012fz,Cline:2013pca,Kahlhoefer:2013dca,Fan:2013yva}
and inspire further simulation of the galactic dynamics in cases where
there are two components of dark matter with naturally very different
mass scales and different self-interaction rates.  Such simulations
would be very helpful to better understand the observational limits on
these theories.  For clusters, another important issue is the fact
that the dark matter may have enough energy to scatter inelastically,
bringing the details of the dark composite sector to the forefront of
the physics; further work is needed to better understand the
implications. We have also pointed out that our models have rich
implications for the early growth of supermassive black holes. The
mechanism by which $\sim 10^9 \, {\rm M}_\odot$ quasars are assembled
as early as redshifts of 6--7 is a mystery, and self-interacting dark
matter could have a major role to play in this story.

If the hidden and visible sectors are closely connected such that the
temperatures remain comparable even at late times, the hidden
glueballs will generically overclose the Universe.  We considered a
depletion mechanism into right-handed neutrinos and found that it can
efficiently remove hidden gluons before confinement.  Self-interaction
strengths required to explain the astrophysical puzzles on small
scales are obtained for glueballino masses $\gtrsim 1$~TeV and
confinement scales $\sim$~MeV.

The possibility of strong self interactions in the dark sector is well
motivated by observations of lower-than-expected dark matter densities
in the centers of galaxies. A strongly interacting hidden sector
naturally realizes this possibility. Even in the simple models
explored in this paper, we have discovered new features that must be
incorporated into numerical simulations to correctly predict the
spatial distribution of dark matter in the central parts of structures
from dwarf galaxies to clusters of galaxies.

\section*{Note Added}

After publishing this article, we became aware of the fact that the glueball and glueballino scattering cross sections are proportional to $1/N^2$ for large $N$.
The plots present an accurate estimate of the scattering cross sections for small $N$, but the cross sections have been overestimated for large $N$.

\begin{acknowledgments}
We are grateful for helpful conversations with David B. Kaplan, Jared
Kaplan, Matthew Reece, Ira Rothstein, Yael Shadmi, Jessie Shelton,
Sean Tulin, and Hai-Bo Yu.  K.B.~thanks the University of California, Irvine, Department of
Physics and Astronomy for their hospitality throughout this work.  K.B.~is
supported in part by the U.S.~DOE Award No.~DE--FG02--92ER40701 and by the
Gordon and Betty Moore Foundation through Grant No.~776 to the Caltech
Moore Center for Theoretical Cosmology and Physics.  J.L.F. and
T.M.P.T. are supported in part by U.S.~NSF Grant No.~PHY--1316792.
T.M.P.T. is further supported in part by the University of California, Irvine through a Chancellor's Fellowship.
M.K. is supported in part by NSF Grants No.~PHY--1214648 and
No.~PHY--1316792. We used the LSODA software from
LLNL~\cite{Hindmarsh:1983ode,Radhakrishnan:1993ode} to solve the
Schr\"{o}dinger equation.  All other calculations were performed with
\textit{Mathematica} 8.0.
\end{acknowledgments}

\bibliography{bibsidm}

\end{document}